\documentclass[aps,prb,twocolumn,preprintnumbers,amsmath,amssymb,
superscriptaddress,citeautoscript,floatfix]{revtex4}

\usepackage{graphicx}
\usepackage{dcolumn}
\usepackage{amsmath}
\usepackage{amssymb}
\usepackage{color}
\usepackage{multirow}

\newcommand{\mSR}{$\mu$SR}
\newcommand{\lem}{LE-$\mu$SR}
\newcommand{\PSI}{Paul Scherrer Institute}

\usepackage[utf8]{inputenc}
\usepackage{siunitx} 
    \graphicspath{{}}
\usepackage{float} 

\usepackage[hidelinks]{hyperref}
\hypersetup{
colorlinks=false,    
linkcolor=red,
citecolor=green,
urlcolor=blue,
pdfborder={0 0 0},
bookmarksopen=true,
bookmarksopenlevel=2,
pdfpagemode=UseOutlines,
pdfstartview={Fit},
pdftitle={Spectroscopic perspective on the interplay between electronic and
magnetic properties of magnetically doped topological insulators},
pdfauthor={Jonas A. Krieger},
pdfsubject={Spectroscopic perspective on the interplay between electronic and
magnetic properties of magnetically doped topological insulators},
pdfkeywords={Low energy muons, SX-ARPES, Topological insulator, magnetic 
doping},
pdfhighlight= /I
}

\begin{document}

\title {Spectroscopic perspective on the interplay between electronic and
magnetic properties of magnetically doped topological insulators }

\author{J.~A.~Krieger}
\affiliation{Laboratory for Muon Spin Spectroscopy, Paul Scherrer
Institute, CH-5232 Villigen PSI, Switzerland}
\affiliation{Laboratorium für Festkörperphysik, ETH-Hönggerberg, CH-8093
Zürich, Switzerland}
\author{Cui-Zu~Chang}
\affiliation{Francis Bitter Magnet Lab, Massachusetts Institute of
Technology, Cambridge, Massachusetts 02139, USA}
\affiliation{Department of Physics, The Penn State
University, University Park, Pennsylvania 16802}
\author{M.-A.~Husanu}
\affiliation{Swiss Light Source, Paul Scherrer Institute, CH-5232 Villigen
PSI, Switzerland}
\affiliation{National Institute of Materials Physics, Atomistilor 405A,
077125 Magurele, Romania}
\author{D.~Sostina}
\affiliation{Swiss Light Source, Paul Scherrer Institute, CH-5232 Villigen
PSI, Switzerland}
\author{A.~Ernst}
\affiliation{Institut f\"ur Theoretische Physik, Johannes Kepler
Universit\"at, A 4040 Linz, Austria}
\affiliation{Max-Planck-Institut f\"ur Mikrostrukturphysik, Weinberg 2,
06120 Halle, Germany}
\author{M.\,M.~Otrokov}
\affiliation{Departamento de F\'{\i}sica de Materiales UPV/EHU, Centro de
F\'{\i}sica de Materiales CFM - MPC and Centro Mixto CSIC-UPV/EHU, 20080
San Sebasti\'an/Donostia, Spain}
\affiliation{Tomsk State University, pr. Lenina 36, 634050 Tomsk, Russia}
\author{T.~Prokscha}
\affiliation{Laboratory for Muon Spin Spectroscopy, Paul Scherrer
Institute, CH-5232 Villigen PSI, Switzerland}
\author{T.~Schmitt}
\affiliation{Swiss Light Source, Paul Scherrer Institute, CH-5232 Villigen
PSI, Switzerland}
\author{A.~Suter}
\affiliation{Laboratory for Muon Spin Spectroscopy, Paul Scherrer
Institute, CH-5232 Villigen PSI, Switzerland}
\author{M.\,G.~Vergniory}
\affiliation{Donostia International Physics Center, P. Manuel de Lardizabal
4, San Sebasti\'an, 20018 Basque Country, Spain}
\affiliation{Department of Applied Physics II, Faculty of Science and
Technology, University of the Basque Country UPV/EHU, Apdo. 644, 48080
Bilbao, Spain}
\author{E.\,V.~Chulkov}
\affiliation{Donostia International Physics Center, P. Manuel de Lardizabal
4, San Sebasti\'an, 20018 Basque Country, Spain}
\affiliation{Departamento de F\'{\i}sica de Materiales UPV/EHU, Centro de
F\'{\i}sica de Materiales CFM - MPC and Centro Mixto CSIC-UPV/EHU, 20080
San Sebasti\'an/Donostia, Spain}
\affiliation{Saint Petersburg State University, 198504 Saint Petersburg,
Russia}
\author{J.~S.~Moodera}
\affiliation{Francis Bitter Magnet Lab, Massachusetts Institute of
Technology, Cambridge, Massachusetts 02139, USA}
\affiliation{Department of Physics, Massachusetts Institute of Technology,
Cambridge, Massachusetts 02139, USA}
\author{V.~N.~Strocov}
\affiliation{Swiss Light Source, Paul Scherrer Institute, CH-5232 Villigen
PSI, Switzerland}
\author{Z.~Salman}
\email[Correspondnig author: ]{zaher.salman@psi.ch}
\affiliation{Laboratory for Muon Spin Spectroscopy, Paul Scherrer
Institute, CH-5232 Villigen PSI, Switzerland}

\date{\today}

\begin{abstract}
  We combine low energy muon spin rotation (LE-$\mu$SR) and soft-X-ray
  angle-resolved photoemission spectroscopy (SX-ARPES) to study the
  magnetic and electronic properties of magnetically doped topological
  insulators, (Bi,Sb)$_{2}$Te$_3$. We find that one achieves a
  full magnetic volume fraction in samples of
  (V/Cr)$_{x}$(Bi,Sb)$_{2-x}$Te$_3$ at doping levels $x \gtrsim
  0.16$. The observed magnetic transition is not sharp in temperature
  indicating a gradual magnetic ordering. We find that the evolution
  of magnetic ordering is consistent with formation of ferromagnetic
  islands which increase in number and/or volume with decreasing
  temperature. Resonant ARPES at the V $L_3$ edge reveals a
  nondispersing impurity band close to the Fermi level as well as V
  weight integrated into the host band structure.  Calculations within
  the coherent potential approximation of the V contribution to the
  spectral function confirm that this impurity band is caused by V in
  substitutional sites.  The implications of our results on the observation of 
the
  quantum anomalous Hall effect at mK temperatures are discussed.
\end{abstract}

\maketitle
\section{Introduction}\label{sec:Intro}
A topological insulator (TI) is a bulk insulator exhibiting an
inverted band structure~\cite{Qi2011}. This, together with time reversal
symmetry (TRS), leads to the presence of a spin polarized Dirac cone
in the surface states which is topologically protected against small,
TRS invariant perturbations. However, TRS breaking in this system,
e.g.~by introducing long range ferromagnetism, opens an energy gap at
the Dirac point. Such a gapped TI has been proposed to exhibit various
new quantum states of matter including the quantum anomalous Hall
(QAH) effect~\cite{Qi2006,Yu2010}, charge induced magnetic mirror
monopoles~\cite{Qi2009} and Majorana excitations when in proximity to
an s-wave superconductor~\cite{Qi2010}.

Recently the QAH effect has been experimentally observed in Cr and V
doped (Bi,Sb)$_2$Te$_3$~\cite{chang2013,chang2015}. This has been
possible only at ultra low temperature (tens of mK), which was
attributed to the presence of bulk valence bands at the binding energy of the
Dirac point~\cite{Li2016}. In addition it has been pointed out that
increasing the homogeneity of the ferromagnetic order might be crucial
to elevate the temperature at which the QAH effect can be
observed~\cite{Feng2016}. Understanding the origin of the magnetism
and the resulting splitting of the topological surface states in
magnetically doped TIs is a prerequisite to tuning and controlling
it. Nevertheless, this issue remains under intense discussion,
particularly due to the lack of suitable methods to directly probe the
magnetic properties of thin films and determine the nature of
magnetism in these
systems~\cite{Vergniory2014,Li2015,Grauer2015,Lachman2015,Ye2015}.

Theoretically, it was predicted that exchange coupling between the
magnetic dopants and the Dirac electrons supports a strong
out-of-plane single ion
anisotropy~\cite{Yu2010,Henk2012,Henk2012a,Nunez2012}. In contrast, in
Cr doped Sb$_2$Te$_3$ it was found that the magnetism at the surface
is oriented in plane, which does not allow for the opening of a
gap~\cite{Yang2013}. Moreover, in Mn doped Bi$_2$Se$_3$ the doping
induced gap in the surface state was shown to be independent of the
magnetism in the sample~\cite{Sanchez2016}.

In order to realize the QAH effect it is essential that the Fermi
level ($E_{\mathrm F}$) is located within the exchange gap of the
topological surface states. In principle, this excludes any type of
free carrier (at $E_{\mathrm F}$) mediated coupling between the
dopants at low temperature. Instead, it has been proposed that the
band inversion in a TI leads to an unusually high spin susceptibility
of the parent compound which allows mediating a ferromagnetic coupling
between the dopants (so called van-Vleck
ferromagnetism)~\cite{Yu2010}. Indeed, a temperature dependent shift
of the V $L_2$~and $L_3$~edges has been measured with electron energy
loss spectroscopy and was attributed to a van-Vleck contribution of
the V core levels to the ferromagnetism~\cite{Li2015}.

More recently, transport and scanning nanoSQUID measurements have
shown that the magnetism in magnetically doped TIs may be attributed to
superparamagnetic-like behavior~\cite{Grauer2015,Lachman2015}. In
addition, a high V density of states near $E_{\mathrm F}$ has been
found using resonant (angular integrated) photoemission spectroscopy
(AIPES) at the V $L_3$~edge and scanning tunneling
spectroscopy~\cite{Peixoto2016,Sessi2016}. This is in agreement with
theoretically predicted nonzero V partial density of states (V-DOS) at
$E_{\mathrm F}$.~\cite{Larson2008,Yu2010}

Several of the above effects potentially hinder the observation of the
QAH effect at higher
temperature~\cite{Li2016,Feng2016,Sanchez2016,Grauer2015,Lachman2015,
  Peixoto2016,Sessi2016} and a deeper understanding of the interplay
between the electronic and magnetic degrees of freedom in these
materials is necessary in order to increase this temperature.

Here we investigate the local magnetic and electronic properties of
thin films of V and Cr doped (Bi,Sb)$_2$Te$_3$ using low energy muon
spin rotation (\lem) and soft X-ray angle resolved photoemission
spectroscopy (ARPES). We show that the ferromagnetic transition is
gradual and strongly depends on the doping level. The evolution of the
magnetic ordering is consistent with magnetic clusters formed at the
transition temperature within a predominantly paramagnetic sample. The
volume fraction of these clusters increases gradually with decreasing
temperature. Our results show that samples of
V$_x$(Bi,Sb)$_{2-x}$Te$_3$ with doping levels $x \gtrsim 0.16$ become
fully magnetic, while at lower doping part of the sample does not
exhibit magnetic ordering. By using resonant soft X-ray ARPES at the V
$L_3$~edge we confirm the presence of a non-dispersing V impurity band
at $E_{\mathrm F}$ as well as spectral weight of V integrated in the
host band structure at lower binding energies ($E_{\mathrm b}$).  We
show that the former is reproduced by density functional theory (DFT)
in the coherent potential approximation (CPA) assuming substitutional
V.  In contrast, if the V is placed in the van-der-Waals (vdW) gap, no
such impuritiy band is formed.  The obtained \lem\ and ARPES results
are used to understand the correlation between the electronic and
magnetic properties of the magnetically doped TIs.

\section{Experiment}\label{sec:Exp}
Films of magnetically doped (Bi,Sb)$_2$Te$_3$ were grown using
molecular beam epitaxy on sapphire~(0001) substrates. The magnetic
dopants were introduced via coevaporation as described in
Refs.~{[\onlinecite{Zhang2011,chang2013b}]} to produce magnetic
M$_x$(Bi$_{1-y}$Sb$_y$)$_{2-x}$Te$_3$, where M is either V or
Cr. Hereafter we refer to the composition of the magnetic TI layer by
M$_{x}$.  The samples investigated by \mSR\ or ARPES were
20~or~10~quintuple layers~(QL) thick, respectively.  Before taking the
samples out of the growth chamber they were capped with a
$3$~or~\SI{5}{nm} thick Te protective layer~\cite{hoefer2015}.  A list
of all the studied samples with relevant details is given in
Table~\ref{tab:Samples}. Note that similarly prepared films exhibit
the QAH effect~\cite{chang2013,chang2015}.

\begin{table}[ht]
    \centering
    \begin{tabular}{|l|l|lr|l|}\hline
     Label & Cap& \multicolumn{2}{c|}{Topological insulator} &
     Technique\\ \hline\hline
      & 5nm Te & 20 QL & Sb$_{2}$Te$_{3}$ &   \lem\\
     V$_{0.13}$ & 5nm Te & 20 QL & V$_{0.13}$Sb$_{1.87}$Te$_{3}$   &  
     \lem\\
     Cr$_{0.2}$ & 5nm Te & 20 QL & Cr$_{0.2}$Sb$_{1.8}$Te$_{3}$  &  
     \lem\\
     V$_{0.08}$ & 3nm Te & 20 QL &
     V$_{0.08}$(Bi$_{0.32}$Sb$_{0.68}$)$_{1.92}$Te$_{3}$ &   \lem\\
     V$_{0.16}$ & 3nm Te & 20 QL &
     V$_{0.16}$(Bi$_{0.32}$Sb$_{0.68}$)$_{1.84}$Te$_{3}$ &   \lem\\
     V$_{0.19}$ & 3nm Te & 20 QL &
     V$_{0.19}$(Bi$_{0.33}$Sb$_{0.67}$)$_{1.81}$Te$_{3}$ &   \lem\\
     V$_{0.23}$ & 3nm Te & 20 QL &
     V$_{0.23}$(Bi$_{0.32}$Sb$_{0.68}$)$_{1.77}$Te$_{3}$ &   \lem\\
      & 3nm Te & 10 QL & (Bi$_{0.31}$Sb$_{0.79}$)$_{2}$Te$_{3}$  
     &      SXARPES\\
     V$_{0.06}$ & 3nm Te & 10 QL & 
     V$_{0.06}$(Bi$_{0.33}$Sb$_{0.67}$)$_{1.94}$Te$_{3}$  &   SXARPES\\
     V$_{0.12}$ &  3nm Te & 10 QL &
     V$_{0.12}$(Bi$_{0.33}$Sb$_{0.67}$)$_{1.88}$Te$_{3}$ &   SXARPES\\
     \hline\hline
     \end{tabular}
     \caption{ List of all the samples and the experimental technique
       which has been used to study them. }\label{tab:Samples}
\end{table}

\lem\ experiments were performed at the $\mu$E4 beam line of the
\PSI~\cite{prokscha2008}. The samples were glued using silver paint
onto Ni-coated sample plates and mounted on the cold finger of a
helium flow cryostat. In these experiments, fully spin polarized muons
with initial polarization chosen to be parallel to the surface of the
sample, were implanted at a tunable energy in the range of
$E=$\SIrange{1}{12}{keV}, corresponding to mean implantation depths of
\SIrange{10}{70}{nm}. This is a key feature which makes \lem\ suited
for studying thin film samples. In contrast, the implantation depth of
surface muons (conventional \mSR) or muons from decay channels
(measurements under high pressure) is on the order of several hundreds
$\mu$m~to~cm, respectively~\cite{Khasanov2016,yaouanc2011}, and
therefore cannot be used to investigate thin samples.

\begin{figure}[tb]
  \includegraphics[width=1.\linewidth]{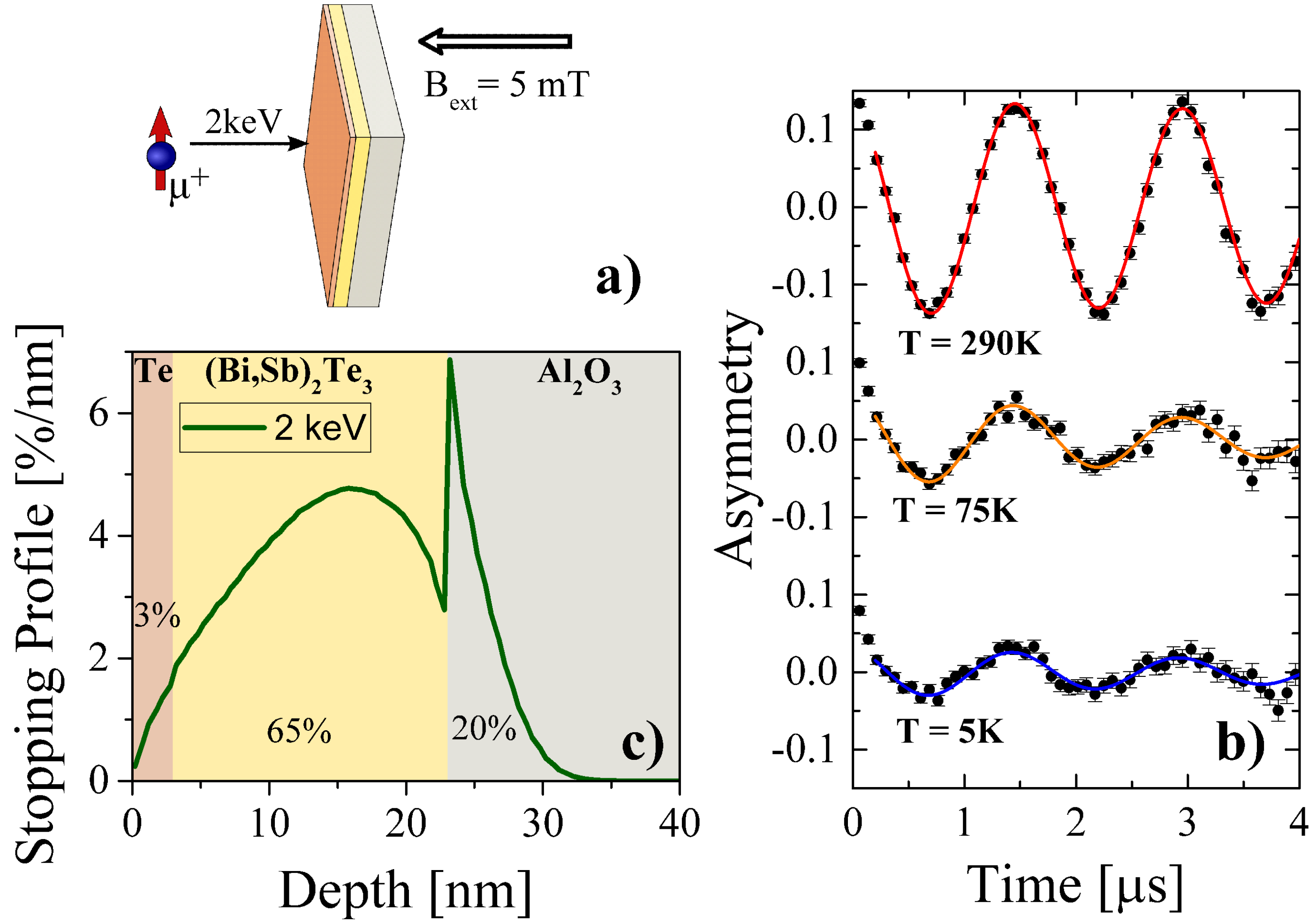}
  \caption{\textbf{a)} Schematic of the experimental configuration,
    where spin polarized muons impinge on the sample with an
    implantation energy $E$. An external magnetic field of \SI{5}{mT}
    is applied perpendicular to the initial muon spin and the sample
    surface. \textbf{b)} Asymmetry spectra measured in the V$_{0.19}$ sample
    with an implantation energy of $E=$\SI{2}{keV}, at different
    temperatures. The solid lines are fits to
    Equation~\ref{eq:AsyExp}.  \textbf{c)} Simulated stopping profile
    of the implanted muons as a function of depth at $E=$\SI{2}{keV}, where 
almost \SI{65}{\percent} of all
    incoming muons stop in the TI layer.}
  \label{fig:Asy}
\end{figure}

The implanted muon decays (lifetime $\tau_{\mu}=\SI{2.2}{\micro s}$)
by emitting a positron preferentially along the muon spin direction.
Therefore, the time evolution of the ensemble average of the muon spin
polarization can be reconstructed by measuring the spatial
distribution of emitted positrons using four detectors around the
sample~\cite{yaouanc2011}. The asymmetry in positron counts between
opposite detectors, $A(t)$, is propotional to the muon spin
polarization along this direction, and reflects the temporal and
spatial distribution of the magnetic fields at the muon stopping site.
Weak transverse field \mSR\ (wTF-\mSR) measurements as a function of
temperature have been performed by field cooling in an applied field
of~\SI{5}{mT} perpendicular to the surface of the sample and the
initial muon spin polarization. The schematic experimental setup is
shown in Figure~\hyperref[fig:Asy]{\ref*{fig:Asy}a}.

Soft X-ray ARPES measurements were conducted at the ADRESS beam line
of the Swiss Light Source at the \PSI~\cite{Strocov2010} using
circularly polarized photons in the energy range of
\SIrange{320}{950}{eV} with a total energy resolution ranging
from~\SI{40}{meV} to~\SI{160}{meV}. At energies corresponding to
resonant photoemission at the V $L_3$~edge the combined beamline and
analyzer resolution was~\SI{60}{meV}.  The experimental geometry
provides a grazing incidence of the X-rays on the sampe as described
in Ref.~[\onlinecite{Strocov2014}].  The measurements were carried out
at \SI{12}{K} to quench the thermal effects reducing the coherent
$\mathbf{k}$-resolved spectral component at high photoelectron
energies~\cite{Braun2013}.  During the experiment the pressure was
kept below \SI{e-10}{mbar}.  In order to correct the angular
dependence of the analyser transmission, all spectra have been
normalized to the diffuse intensity measured above $E_{\mathrm F}$.
X-ray absorption spectroscopy (XAS) measurements were also performed
on the same beamline by measuring the total electron yield (TEY) via
the sample drain current normalized by the incoming photon intensity.

For the ARPES measurements, the Te capping was first partially removed
by sputtering the sample in \SI{5e-6}{mbar} Ar at \SI{0.25}{keV} for
\SI{20}{min}. The remaining cap layer was evaporated by quickly heating
the sample up to \SI{340}{^\circ C} where it was kept for
\SI{10}{min}. Then it was cooled down to \SI{300}{^\circ C} where it
was kept for \SI{20}{min} to prevent readsorption of the evaporated
Te, before cooling it to cryogenic temperatures and performing the
ARPES measurements.  During this process, the temperature of the
samples was monitored using an optical pyrometer and the pressure in
the preparation chamber was kept below \SI{e-7}{mbar}.

\section{Results}\label{sec:Results}
\subsection{Low energy muon spin rotation}\label{sec:LEM}
Representative wTF-\mSR\ asymmetry spectra, measured at \SI{2}{keV}
implantation energy and different temperatures are presented in
Fig.~\hyperref[fig:Asy]{\ref*{fig:Asy}b}.  At high temperatures an
almost undamped, large amplitude precession signal was observed. In
contrast, a much smaller amplitude with a larger damping rate was
measured at low temperatures, where a considerable part of the
asymmetry decayed at very early times.  At the implantation energy
used in these measurements, most of the muons stop inside the TI
layer, while less than \SI{3}{\percent} stop inside the Te capping and
less than \SI{20}{\percent} in the sapphire substrate.
Figure~\hyperref[fig:Asy]{\ref*{fig:Asy}c} shows the simulated
stopping distribution profiles of the muons at the corresponding implantation 
energy, which was calculated using the \texttt{TRIM.SP} Monte-Carlo
code~\cite{morenzoni2002}.  Measurements as a function of implantation
energy confirmed the validity of this simulation, see Figure~1 in the
supporting information~(SI)~\cite{SI}.

In a transverse field measurement where the applied field $B_{0}$ is
much larger than the internal static fields sensed by the muons, one
expects to measure an asymmetry oscillating at the Larmor frequency,
$\omega_{\mathrm L} = \gamma_\mu B_0$, where
$\gamma_\mu = 2 \pi \times \SI{135.5}{MHz/T}$ is the muon gyromagnetic
ratio. However, when the internal static field is much larger than the
applied field, as is typically the case below a magnetic transition,
the oscillating asymmetry becomes heavily damped. Due to the finite
temporal resolution of the \mSR\ technique this can lead to an
effective loss of initial asymmetry. Therefore, muons stopping in a
magnetically ordered part of the sample or in the nickel sample
holder~\cite{Saadaoui2012} experience a larger static magnetic field
and appear as a missing fraction in the asymmetry spectra. On the
other hand, muons stopping in a paramagnetic region of the sample will
precess at~$\sim \omega_{\mathrm L}$. Hence, considering only the
oscillating signal after \SI{0.2}{\micro s} allows to estimate the
paramagnetic volume fraction of the sample, while the missing
asymmetry at early times corresponds to the magnetic volume fraction.

\begin{figure}[htb]
  \includegraphics[width=1.\linewidth]{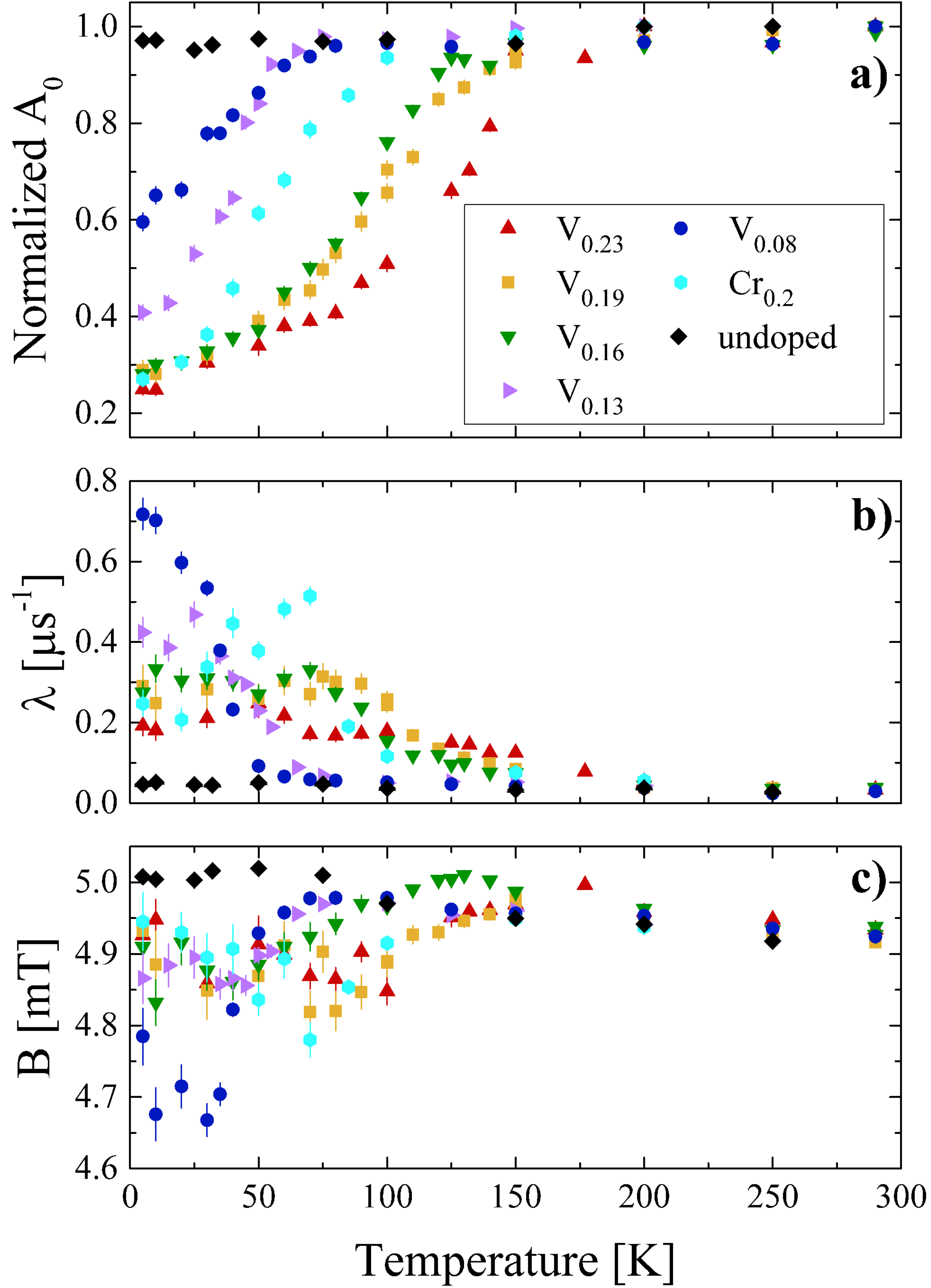}
  \caption{Temperature dependence of \textbf{a)} the normalized $A_0$,
    \textbf{b)} $\lambda$ and \textbf{c)} $B$ obtained from the fits
    for the different M$_x$(Bi$_{1-y}$Sb$_y$)$_{2-x}$Te$_3$ samples
    which are denoted by M$_{x}$.  }
  \label{fig:AsyLamB}
\end{figure}

Following this logic, we used the \texttt{Musrfit}
software~\cite{Suter2012} to fit the measured \mSR\ spectra between
\SI{0.2}{\micro s} and \SI{8}{\micro s} to an exponentially damped
oscillating function,
\begin{equation}\label{eq:AsyExp}
  A(t) = A_0e^{-\lambda t}\cos\left(\gamma_\mu Bt+ \varphi\right),
\end{equation}
where $A_0$ is the extrapolated initial amplitude of the oscillating
asymmetry at $t= \SI{0}{\micro s}$, $B$ is the mean field at the
muons' stopping sites and $\lambda$ is the damping rate, which
corresponds primarily to the width of the static local field
distribution. The phase, $\varphi$, is determined by the initial muon
spin direction and the geometry of the detectors. The temperature
dependence of $A_0$ (normalized to its value in the paramagnetic phase
at 290 K), $B$ and $\lambda$ obtained from the fits of all studied
samples are shown in Figure~\ref{fig:AsyLamB}.

Note that the undoped sample exhibits an almost temperature
independent $A_0$ and $\lambda$, as expected for a non-magnetic
sample. In contrast, we observe a drop in $A_0$ upon cooling the doped
samples, which we associate with the appearance of strong static
fields in portions of the sample. This drop can be attributed to a
magnetic ordering at a transition temperature $T_c$. Clearly, $T_c$
and the relative size of drop in $A_0$ depends strongly on the dopant
and doping level in the sample. We also note that the drop in $A_0$ is
accompanied by an increase in $\lambda$ and a decrease in $B$.

\begin{figure}[tb]
  \includegraphics[width=0.95\linewidth]{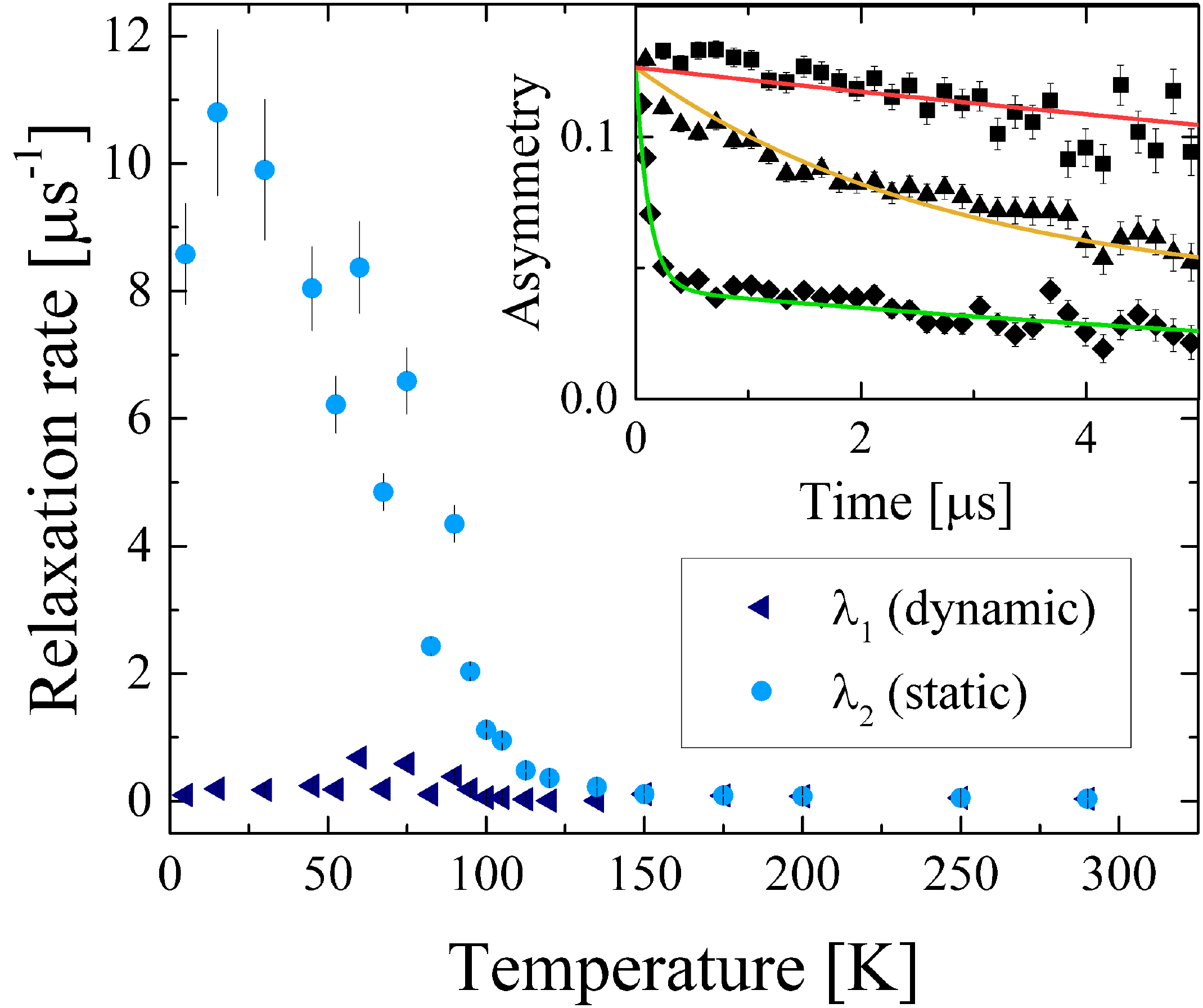}
  \caption{Dynamic ($\lambda_1$) and static ($\lambda_2$) zero field
    relaxation rate as a function of temperature in the V$_{0.19}$
    sample. The inset shows typical asymmetry spectra at temperatures
    of \SI{290}{K}, \SI{120}{K} and \SI{5}{K} (from top to
    bottom). The solid lines are fits to Equation~\ref{eq:ZF}.}
  \label{fig:ZF}
\end{figure}

For comparison, measurements in zero field (ZF) on the V$_{0.19}$
sample are shown in Figure~\ref{fig:ZF}. At high temperature, the
asymmetry follows an exponential-like behavior with a damping rate
wich increases sharply below $T_c$.  We also note a change of the
relaxation from a single exponential-like behavior to a bi-exponential
relaxation below $T_c$, as is typical for systems undergoing magnetic
ordering~\cite{yaouanc2011}.  Therefore, we fit the data to the sum of
a two exponentials,
\begin{equation}\label{eq:ZF}
  A(t) = A_0 \left(\frac{1}{3}e^{-\lambda_1 t}+\frac{2}{3}e^{-\lambda_2
      t}\right).
\end{equation}
This simplified model assumes that below the magnetic transition the
implanted muons experience a Lorentzian distribution of randomly
oriented static fields with an additional small and fluctuating
component.  In this case, an ensemble average of one third of the
initial muon spin polarization is aligned parallel to the randomly
oriented static local field, while two thirds are perpendicular to
it. The 1/3 component can relax only due to fluctuations in the
magnetic field, and hence is referred to as a dynamic component. On
the other hand, the damping of the 2/3 components is due to incoherent
precession and depolarization which is determined by the width of the
static field distribution and is referred to as a static
component~\cite{yaouanc2011}. Note that the direction of the local
field at the muon stopping position strongly depends on the relative
(random) location of the nearest dopant atom. Therefore, the random
orientation of the local dipolar field does not necessarily imply
randomly oriented magnetic moments.

Despite ignoring the presumably different contributions from various
layers of the sample and substrate, the ZF measurements fit very well
to Eq.~(\ref{eq:ZF}), confirming the observed magnetic ordering from
TF measurements and allowing a qualitative understanding of the
behavior of the sample. In particular it shows that the damping rates
above $T_c$ are equal and almost temperature independent, reflecting
the dynamic nature of the muon spin relaxation. Below $T_c$ we observe
a strong increase of the depolarization of $2/3$ component which is
associated with the appearance of static magnetic fields in the sample
due to the magnetic ordering.

\subsection{Soft X-ray ARPES}\label{sec:ARPES}
\begin{figure*}
  \includegraphics[width=1.\linewidth]{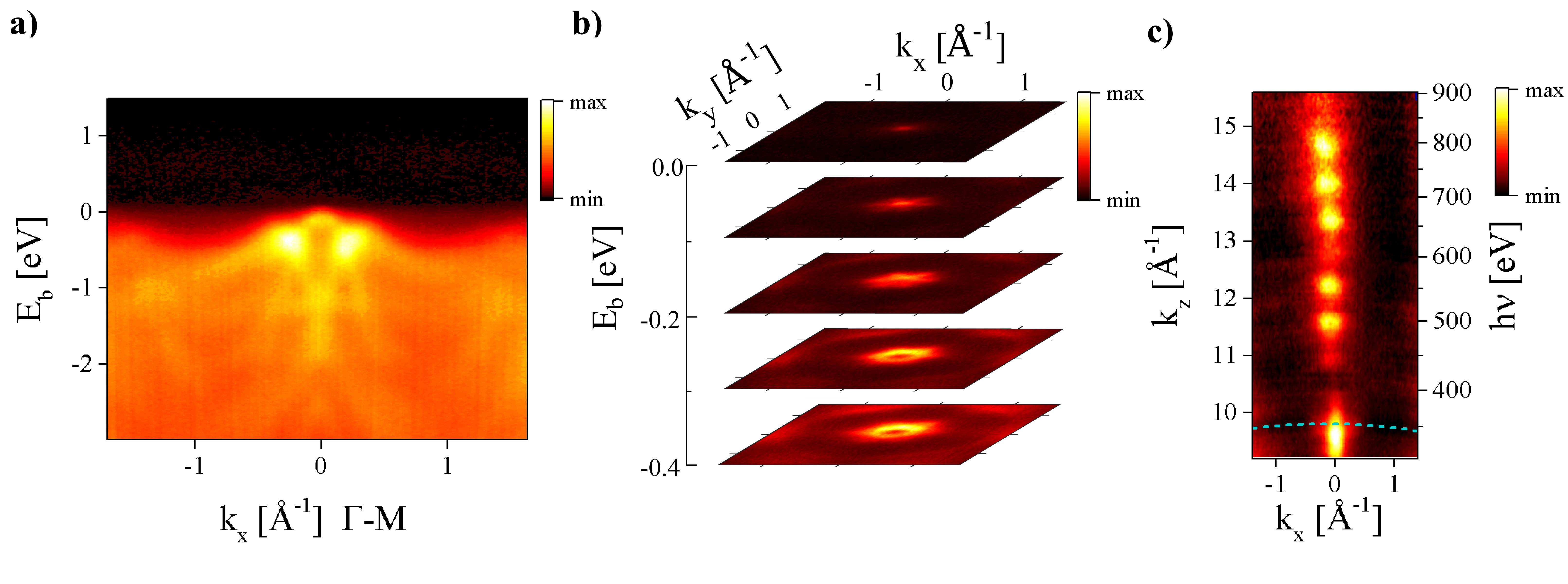}
  \caption{\textbf{a)} Band structure at \SI{355}{eV} of the
    V$_{0.06}$ sample measured along the $\Gamma-M$ direction.
    \textbf{b)} Constant energy cut maps measured at
    $h\nu=\SI{355}{eV}$.  \textbf{c)} Fermi surface map along
    ($k_x$,~$k_z$). The $h\nu$ values at $k_x=0$ are shown in the
    right axis. The dashed line indicates the position of the cut
    shown in a) and b). }
  \label{fig:KNavi}
\end{figure*}

The experimental band structure of the~V$_{0.06}$ sample measured at photon 
energy 
$h\nu=\SI{355}{eV}$ is shown in Figure~\ref{fig:KNavi}.  The constant
$E_{\mathrm b}$ cuts as a function of the $(k_x,k_y)$ coordinates
(Figure~\ref{fig:KNavi}b) exhibit a point-like feature at
$E_{\mathrm F}$, which can be associated with the topological surface
state, and the hexagonally symmetric, flower-shaped bulk valence band
below.  This is similar to what is seen in ultra-violet
ARPES,~cf.~[\onlinecite{Li2016}].  We note that the dispersions are
less sharp than in ultra-violet ARPES measurements on similar
samples~\cite{Chen2009,Li2016}.  This worsens with increasing $h\nu$,
cf.~the measurement at \SI{510}{eV} in~Figure~\ref{fig:DiffPlots}a. We
attribute this effect to the doping induced distortions of the crystal
lattice, which can be considered as ``frozen phonons'' affecting the
coherent ARPES intensity similarly to the thermal effects which occur
at high photon energies~\cite{Braun2013}.

In Figure~\hyperref[fig:KNavi]{\ref*{fig:KNavi}c}, the measurement as
a function of $h\nu$ reveals that the surface state
exhibits intensity modulations without dispersing as a function of
$k_z$. To compensate for the effect of the decrease in the
photoemission cross section with increasing $h\nu$ in
Figure~\hyperref[fig:KNavi]{\ref*{fig:KNavi}c}, each measurement was
normalized to the total AIPES intensity outside the $\Gamma$ point,
within $\pm$\SI{0.1}{eV} around $E_{\mathrm F}$.

\begin{figure}[tb]
  \includegraphics[width=1.\linewidth]{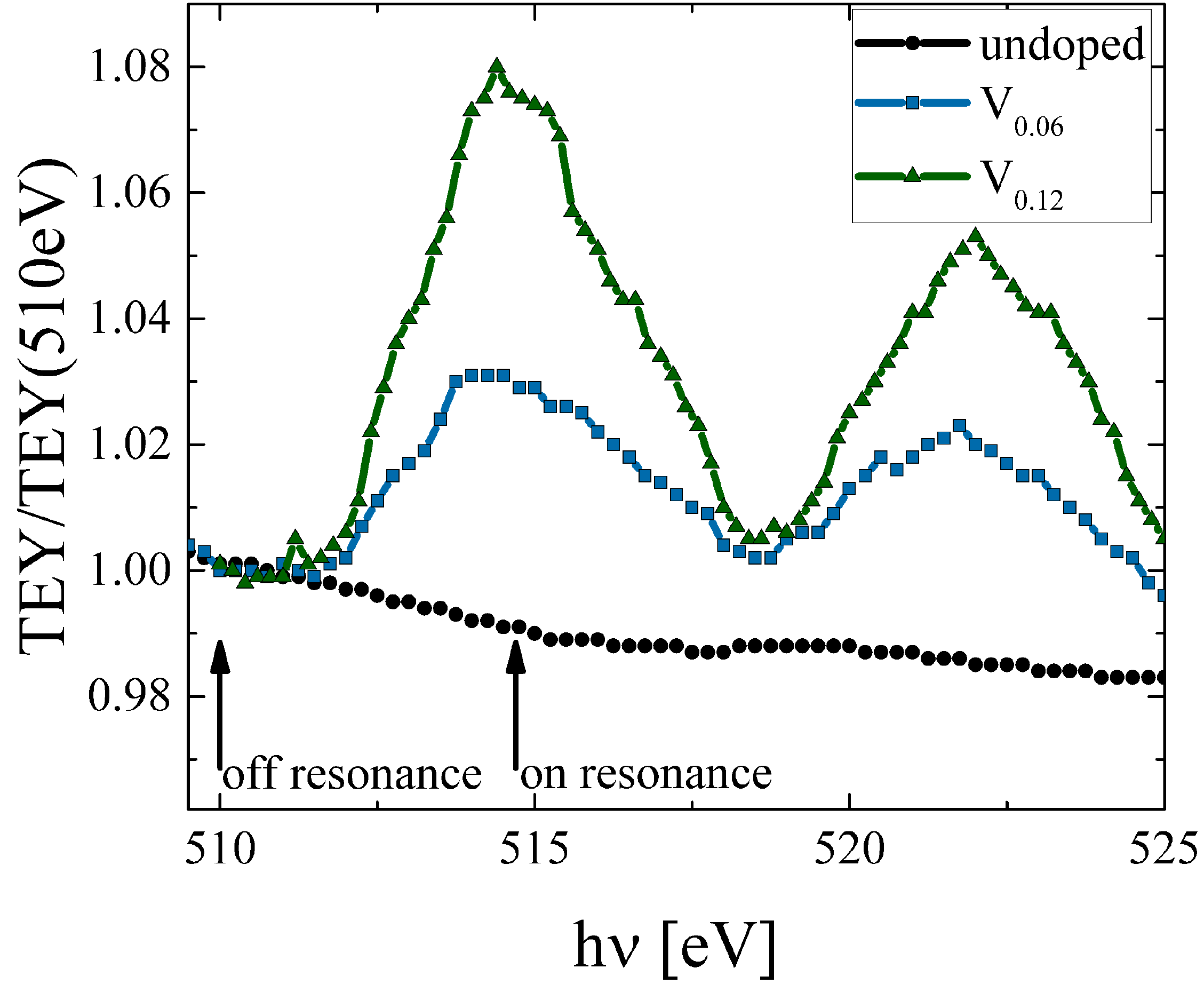}
  \caption{Normalized TEY at the V $L_{2}$~and $L_{3}$~edges for
    different dopings. The arrows indicate the energies where the on-
    and off-resonant ARPES spectra were measured.  }
  \label{fig:XAS}
\end{figure}

Figure~\ref{fig:XAS} shows the XAS across the V $L_{2}$~and
$L_{3}$~edges. The deviation from the theoretical branching ratio of
2:1 and satellites to the peaks have been reported previously in
similar samples and were attributed to final state effects
([\onlinecite{Peixoto2016}]~and references therein).  The peak
positions are consistent with the literature values, however the
peak-width is larger than what is usually observed in these
samples~\cite{Peixoto2016,Sessi2016}. As we will discuss later, this
could be caused by different chemical environments of the V atoms.

\begin{figure*}[t]
  \includegraphics[width=.999\linewidth]{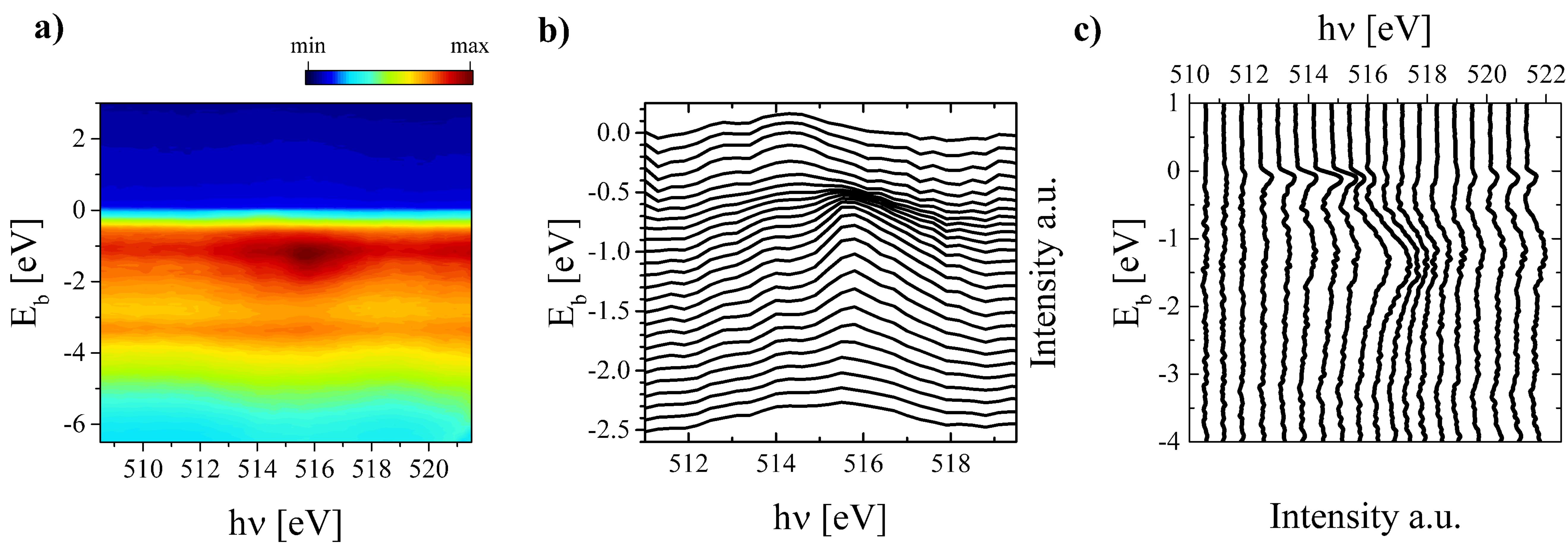}
  \caption{\textbf{a)} Raw AIPES of the V$_{0.06}$ sample as a
    function of $h\nu$ across the $L_{3}$~edge of V.  \textbf{b)}
    Angle integrated constant $E_b$ cuts.  \textbf{c)} The
    difference between AIPES measured at $h\nu$ and off resonance
    (i.e. at \SI{510}{eV}). }
  \label{fig:Vres}
\end{figure*}

In Figure~\ref{fig:Vres} we plot the raw AIPES intensity of the
V$_{0.06}$ sample as a function of $E_{\mathrm b}$ and $h\nu$ where
the latter is scanned across the V $L_{3}$~edge.  The spectra were
integrated along the $\Gamma$-M direction.  There is a clear intensity
increase at the V $L_{3}$-resonance around
$h\nu\approx\SI{515.7}{eV}$, at $E_{\mathrm b}\approx\SI{-1.2}{eV}$.
A careful analysis of constant $E_{\mathrm b}$-cuts, shown in
Figure~\hyperref[fig:Vres]{\ref*{fig:Vres}b}, reveals that at
$h\nu\approx\SI{514.2}{eV}$ there is another resonance close to
$E_{\mathrm F}$.  Note that there is no intensity drop at low $h\nu$
like it is sometimes observed for Fano resonances.  The two resonances
can be distinguished more clearly by looking at the energy
distribution curves (EDC) in
Figure~\hyperref[fig:Vres]{\ref*{fig:Vres}c} from which the
off-resonance contribution measured at \SI{510}{eV} has been
subtracted.

\begin{figure}[tb]
  \includegraphics[width=1.\linewidth]{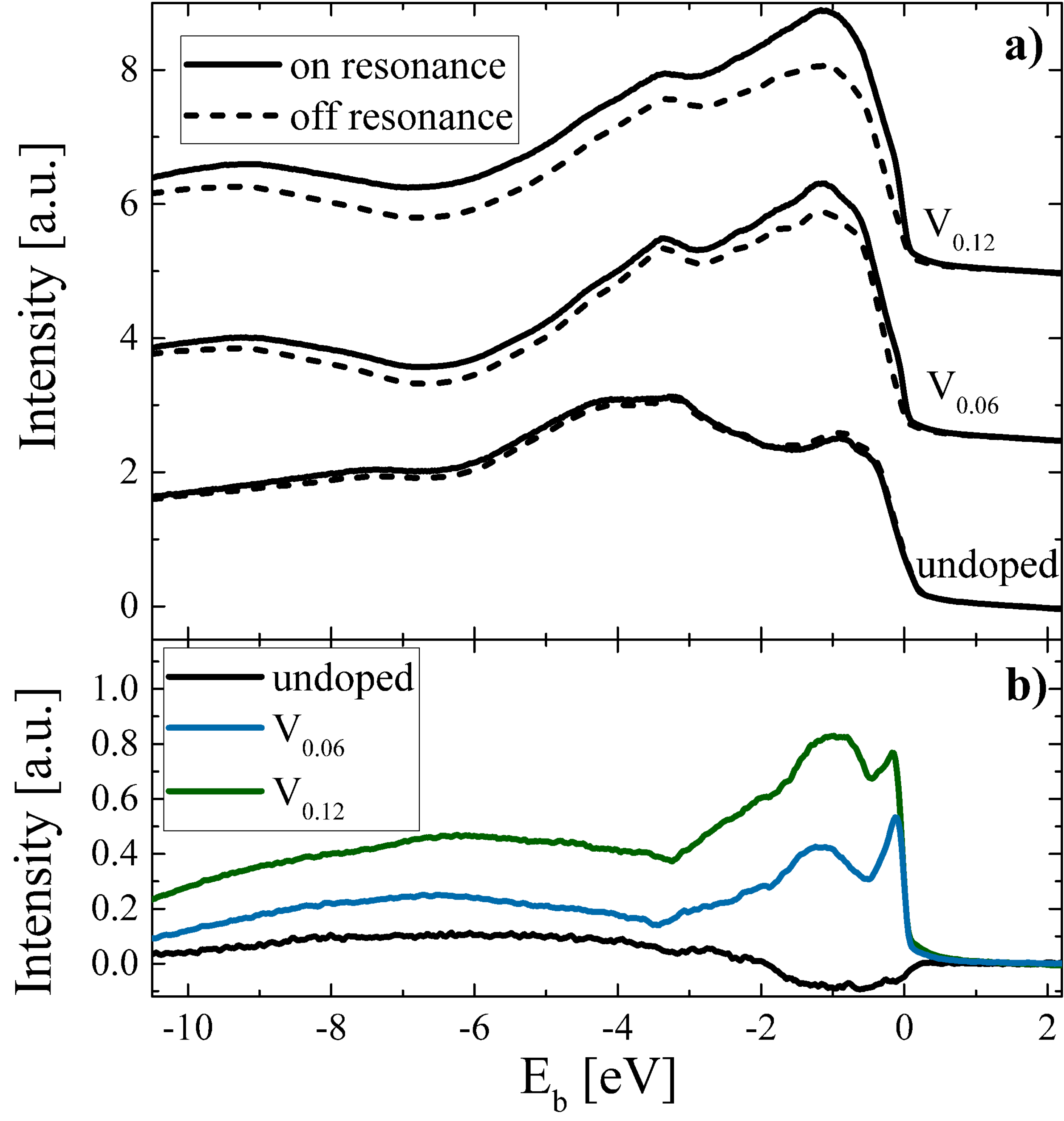}
  \caption{\textbf{a)} Resonant AIPES at different doping
    levels. \textbf{b)} Intensity difference between the on and off
    resonance measurements shown in a). }
\label{fig:ResInt}
\end{figure}

\begin{figure*}[tb]
  \includegraphics[width=1.\linewidth]{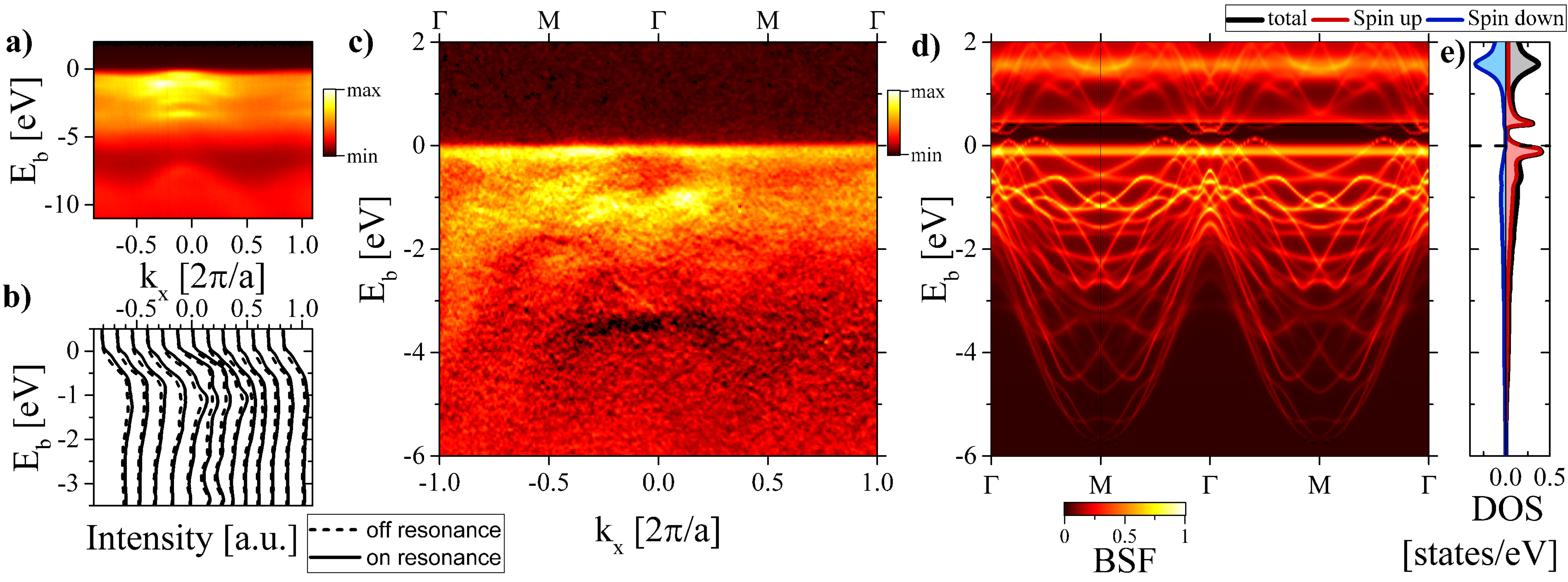}
  \caption{ \textbf{a)} Off resonance ARPES spectrum measured at
    $h\nu=\SI{510}{eV}$.  \textbf{b)} Energy dependent cuts of the on
    and off resonant spectra which were used to calculate c).
    \textbf{c)} Intensity difference of the spectra measured on and
    off resonance for the V$_{0.06}$ sample.  \textbf{d)} Calculated V
    contribution to the BSF of
    V$_{0.06}$(Bi$_{0.33}$Sb$_{0.67}$)$_{1.94}$Te$_3$, assuming that V
    only occupies substitutional positions. \textbf{e)} The
    corresponding spin-resolved V-DOS.  }
  \label{fig:DiffPlots}
\end{figure*}

Resonant ARPES measurements have been performed by tuning the photon
energy to the V $L_{3}$~peak measured in XAS (\SI{514.7}{eV}) as
indicated in Figure~\ref{fig:XAS}. Off resonance reference
measurements were performed at $h\nu=\SI{510}{eV}$.  The corresponding
AIPES at different doping levels are compared in
Figure~\ref{fig:ResInt}.  Although the spectral shape of the undoped
sample might have been affected by its incomplete decapping, it does
not show any resonance at the V $L_3$~edge neither in XAS nor in
resonant AIPES.  The difference of the AIPES measured on and off
resonance, which reflects the
V-DOS~\cite{Molodtsov1997,Kobayashi2014}, is shown in
Figure~\hyperref[fig:ResInt]{\ref*{fig:ResInt}b}.  In the doped
samples the two peaks are located at similar $E_{\mathrm b}$,
suggesting that the behavior observed in Figure~\ref{fig:Vres} is
independent of the doping level.  Note that we can not resolve a shift
of the DOS on the order of \SI{20}{meV} at different doping levels due
to a change in host carrier density that has been previously reported
in Cr-doped (Bi,Sb)$_2$Te$_3$~\cite{Ye2015}.

The EDCs measured on and off resonance (corresponding to the AIPES in
Figure~\ref{fig:ResInt}a) on the V$_{0.06}$ sample are shown in
Figure~\ref{fig:DiffPlots}b.  Subtracting them gives the angle
resolved difference plot in Figure~\ref{fig:DiffPlots}c.  It is
important to point out that the resonant contribution close to
$E_{\mathrm F}$ forms a non-dispersive impurity band, whereas the one
around \SI{-1.2}{eV} seems to be dispersive and hybridized with the
host band structure. We also note that the behavior of the V$_{0.12}$
sample is similar, but more smeared due to the larger disorder.

\subsection{Theory}
Electronic structure calculations were performed by employing density
functional theory within the generalized gradient
approximation.\cite{Perdew1992a} We used a full potenital relativistic
spin-polarized Green function method,\cite{Geilhufe2015} in which
disorder effects were treated within the coherent potential approach
(CPA).\cite{Gyorffy1972} The angular momentum cutoff was set to
$l_{max}=3$. We have used 32 Gaussian quadrature points to carry out a
complex energy contour integration, while for the integration over the
irreducible Brillouin zone we used a 30$\times$30$\times$30 ${\bf k}$
points mesh. The experimental lattice constants and atomic coordinates
of (Bi,Sb)$_2$Te$_3$ were adopted for the
calculations.\cite{Stasova1970} According to the available
experimental data,\cite{Dyck2002, Dyck2003, Dyck2005,
  Kulbachinskii2005, Choi2004, Zhou2006a, Hor2010, Song2012, West2012,
  Kou2012, Schlenk2013, Yee2015, Polyakov2015, Liu2015} 3$d$
transition metal impurities in tetradymite-like chalcogenides
substitute typically cation atoms (Bi and Sb). However, STM studies
also reveal van der Waals positions \cite{Song2012, Yee2015} of Fe
atoms in the bulk Bi$_2$Se$_3$. Based on these data, we considered
that V may occupy the substitutional position in the
(Bi,Sb)-sublattice and the octahedral interstitial site in the vdW
gap.

\begin{figure}[tb]
  \includegraphics[width=1.\linewidth]{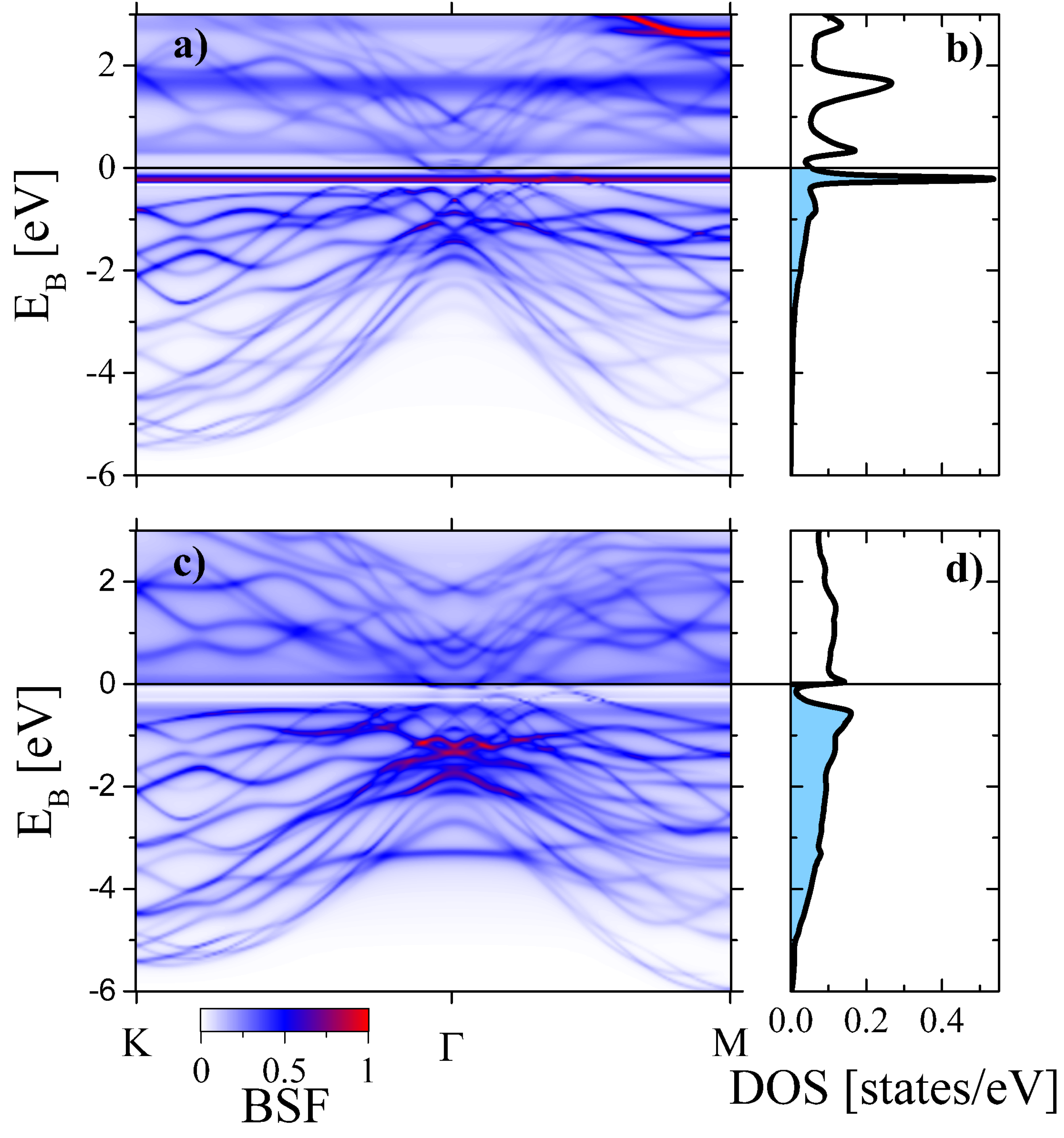}
  \caption{Calculation on
    V$_{0.06+0.03}$(Bi$_{0.33}$Sb$_{0.67}$)$_{1.94}$Te$_3$ with V
    occupying both substitutional and vdW gap positions.  The
    contribution of substitutional V \textbf{a)} and V in the vdW gap
    \textbf{c)} to the BSF, with the corresponding V-DOS in
    \textbf{b), d)}, respectively.}
  \label{fig:DFT}
\end{figure}

To extract the V contribution to the Bloch spectral function (BSF), we
calculated the Fourier transformed site and angular momentum projected
Green function including non-site-diagonal terms within a coherent
potential approximation. In this approach a direct extraction of the V
contribution is not possible because of the non-site-diagonal terms.
However, since in these systems only V atoms have $d$ electrons, the
Bloch spectral function projected into the angular momentum channel
$l=2$ can be associated with the V contribution to the BSF.  The
result for V$_{0.06}$(Bi$_{0.33}$Sb$_{0.67}$)$_{1.94}$Te$_3$
considering only V in substitutional positions is shown in
Figure~\ref{fig:DiffPlots}d.  A second calculation including aditional
V in the vdW gap of
V$_{0.06+0.03}$(Bi$_{0.33}$Sb$_{0.67}$)$_{1.94}$Te$_3$ is shown in
Figure~\ref{fig:DFT}.

\section{Discussion }\label{sec:Disc}
\subsection{Muon spin rotation}
The strong increase of the static ZF relaxation rate below $T_c$
confirms the magnetic origin of the features we observe in wTF-\mSR\
measurements below the same temperature. The magnetic moments of the
dopant atoms, which give only a small, motionally narrowed relaxation
rate above the transition, slow down with decreasing temperature and
produce a broad, static field distribution below $T_c$. In the ZF
measurements, a small increase of the dynamic relaxation is also
observed below the transition, probably due to the slowing down of
magnetic field fluctuations from the dopant moments.  We also point
out that the rate of the static depolarization at low temperature
decreases with increasing doping level (not shown). This indicates a
broader field distribution in the lower doping samples (as expected)
and is generally consistent with the corresponding decrease in the wTF
depolarization rates (Figure~\ref{fig:AsyLamB}b).

\begin{figure}[htb]
  \includegraphics[width=0.95\linewidth]{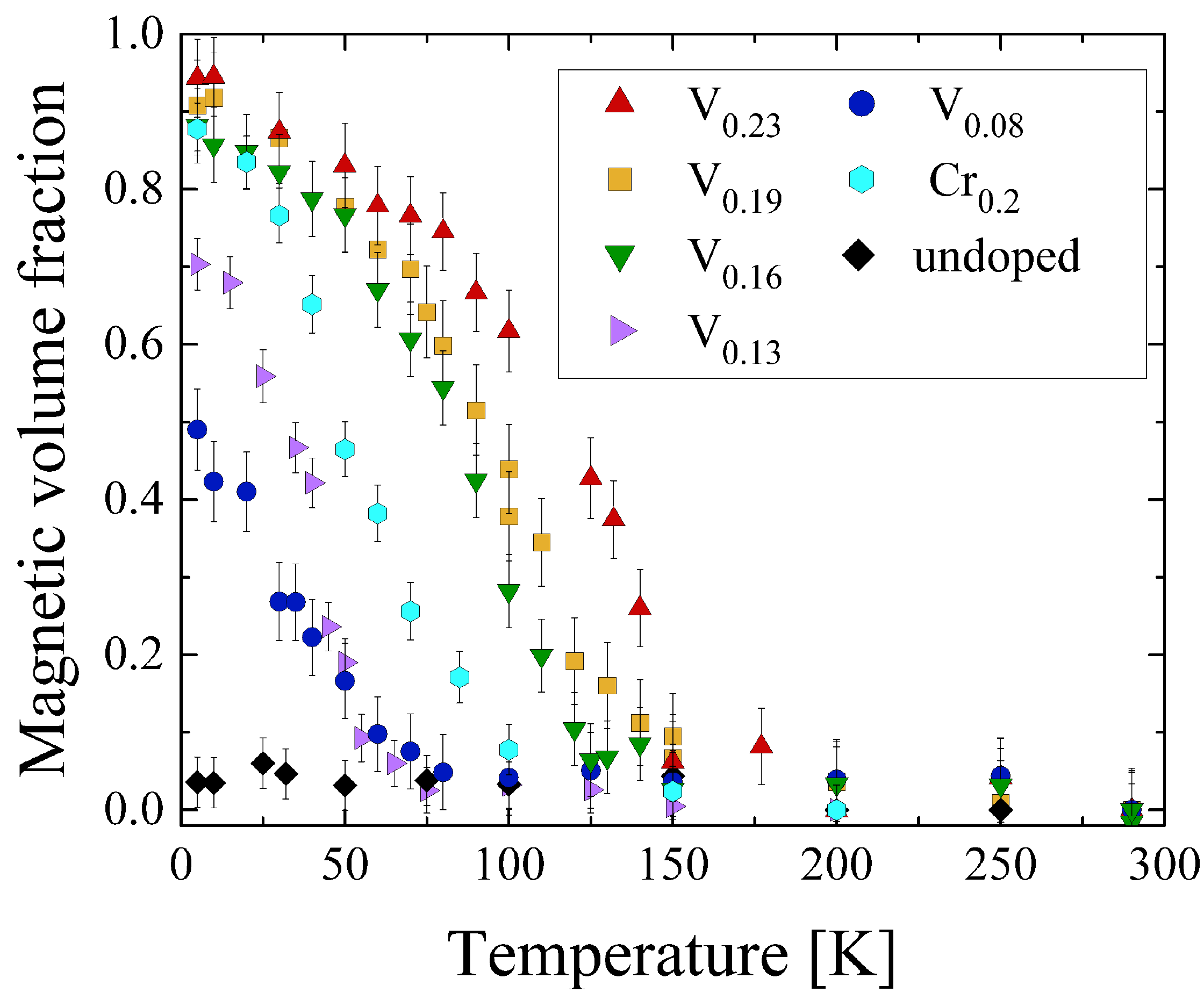}
  \caption{Magnetic volume fraction as a function of temperature,
    measured in the samples listed in Table~\ref{tab:Samples}.}
  \label{fig:volFrac}
\end{figure}

The onset of the magnetic transition at $T_c$, which coincides with
the drop in $A_0$, increases with doping level. We also note that the
transition temperature is generally higher for the V doped samples
than for the Cr doped sample, even at lower doping levels. This is in
agreement with the ferromagnetism that is commonly observed in these
materials~\cite{Dyck2002,Dyck2005,Kulbachinskii2005,Zhou2006a,chang2013,
  chang2015, Ye2015}.  As mentioned earlier, the gradual decrease of
the wTF initial asymmetry below $T_c$ indicates the gradual formation
of ferromagnetic regions, in an otherwise paramagnetic sample.  At
$T_c$, ferromagnetism is established in small islands which increase
in number and/or size as the temperature is decreased.  One important
aspect of the results in Figure~\ref{fig:AsyLamB} is the fact that in
the samples with dopant concentrations $x \gtrsim 0.16$ the initial
asymmetry seems to decrease to the same value at the lowest
temperature, $\sim\SI{5}{K}$. Given the similar geometry of the
measured samples, this is a clear indication that at this temperature
the TI layer is fully magnetic and that the remaining asymmetry is due
to muons stopping in the Te capping layer, the sapphire substrate or
background contribution from muons landing outside the sample.  From these
measurements we can evaluate the magnetic volume fraction of the doped
TI layer by considering the contribution of muons stopping in each
layer to the measured asymmetry, normalized by the full asymmetry
above $T_c$.  Here we assume that muons landing in the non-magnetic Te
capping maintain their full polarization, while those stopping in the
Ni backing depolarize immediately and do not contribute to the
measured asymmetry. Furthermore we assume (based on measurements on
the bare substrate) that muons in sapphire maintain only
\SI{42(7)}{\percent} of their polarization due to Muonium formation.
The full details of these estimates and the reference measurements are
given in the SI~\cite{SI}. The resulting volume fraction as a function
of temperature is shown in Figure~\ref{fig:volFrac}.

\begin{figure}[htb]
  \includegraphics[width=0.95\linewidth]{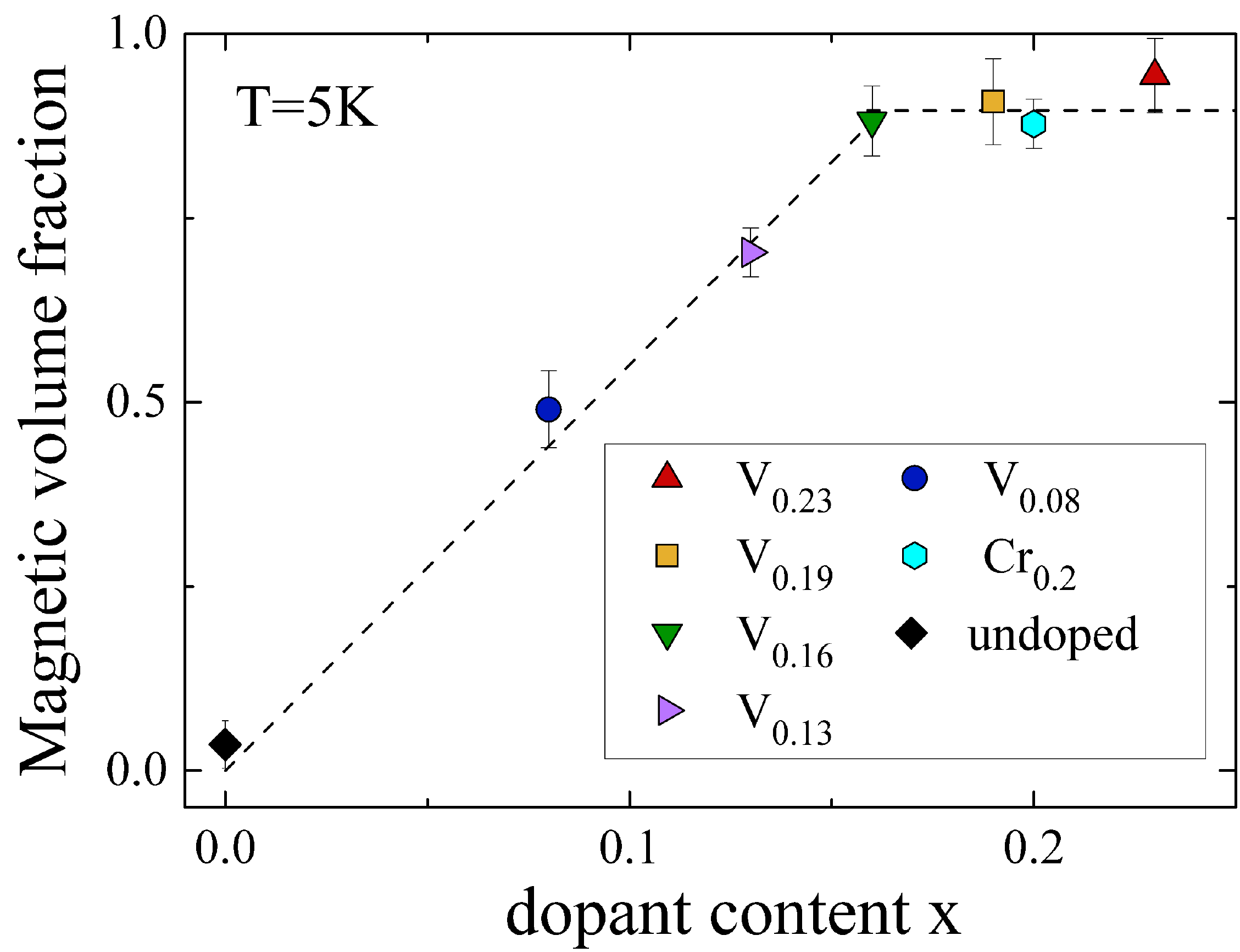}
  \caption{Magnetic volume fraction at the lowest temperature of
    \SI{5}{K} as a function of the dopant concentration~$x$. The
    dashed line is a guide to the eye, showing that the volume
    fraction increases almost linearly with $x$ until it saturates at
    $x\approx 0.16$.}
  \label{fig:volFrac5K}
\end{figure}

These results clearly show that the volume fraction at low temperature
increases with increasing doping level saturating at a doping of
$x\approx0.16$, as shown in Figure~\ref{fig:volFrac5K}. Note that this
saturation value of the magnetic volume fraction is consistent with a
full volume fraction considering the systematic uncertainty in estimating the 
probability of
muonium formation in the sapphire substrate (see full details in the
SI~\cite{SI}). The samples with V doping levels of~$x=0.13$
and~$x=0.08$ do not reach a full magnetic volume fraction at the
lowest measured temperature~(\SI{5}{K}), even when extrapolated to
$T\rightarrow\SI{0}{K}$. This is one of the important aspects of our
results, since such samples, seemingly with only partial magnetic
volume fraction, have been shown to exhibit the QAH at very low (mK)
temperature~\cite{chang2015,Grauer2015}. Remember, the QAH effect can
occur only if the topological surface states are gapped, e.g. by a
static magnetic field perpendicular to the surface of the sample. A
possible explaination for the source of this static field throughout
the sample may be drawn from our measurements. The negative shift of
$B$ (Figure~\ref{fig:AsyLamB}c) implies that even in the paramagnetic
regions of the sample there is on average a net static field
perpendicular to the surface. At low enough temperatures these fields
may be sufficient to open the required gap in the surface states which
is necessary for the observation of the QAH effect.

The drop in $A_0$ is consistent with both, partial volume
ferromagnetism and superparamagnetism as has been reported on similar
samples~\cite{Lachman2015,Grauer2015}. However, clues concerning the
nature of the magnetic ordering in these systems can be obtained from
a careful inspection of the temperature and field dependence of $B$
(Figure~\ref{fig:AsyLamB}c).  In the undoped Sb$_2$Te$_3$ we observe
an increase of $B$ with decreasing temperature, up to \SI{2}{\percent}
of its value at room temperature (RT).  The exact source of this shift
is unknown, but it is most probably of hyperfine origin.  A similar
shift in $B$ is observed above $T_c$ in the magnetically doped
samples.  However, below $T_c$ we detect a negative shift in $B$,
which reaches \SI{5}{\percent} of the RT value. This surprising effect
is most pronounced in the lowest doping sample.  Such large (and
negative) field shifts have been reported in other materials, e.g. in
the diluted magnetic semiconductor
Cd$_{1-x}$Mn$_x$Te~\cite{Golnik1986,Ansaldo1988} and in
MnSi~\cite{Hayano1980JPSJ,Lancaster2016}. These were attributed to a
large hyperfine interaction of the muon with the screening electrons,
which results in a shift proportional to the applied
field~\cite{yaouanc2011,Golnik1986,Ansaldo1988,Hayano1980JPSJ,
  Lancaster2016}.  However, in the V$_{0.08}$ sample we detect a shift
which is not proportional to the applied field (Figure~2 in the
SI~\cite{SI}). In addition field coled and zero field cooled
measurements yield significantly different shifts (Figure~3 in the
SI~\cite{SI}). These observations exclude a shift due to hyperfine
interaction and support a scenario where the shift is due to dipolar
fields from ferromagnetic islands acting as superparamagnets.

\begin{figure}[htb]
  \includegraphics[width=0.9\linewidth]{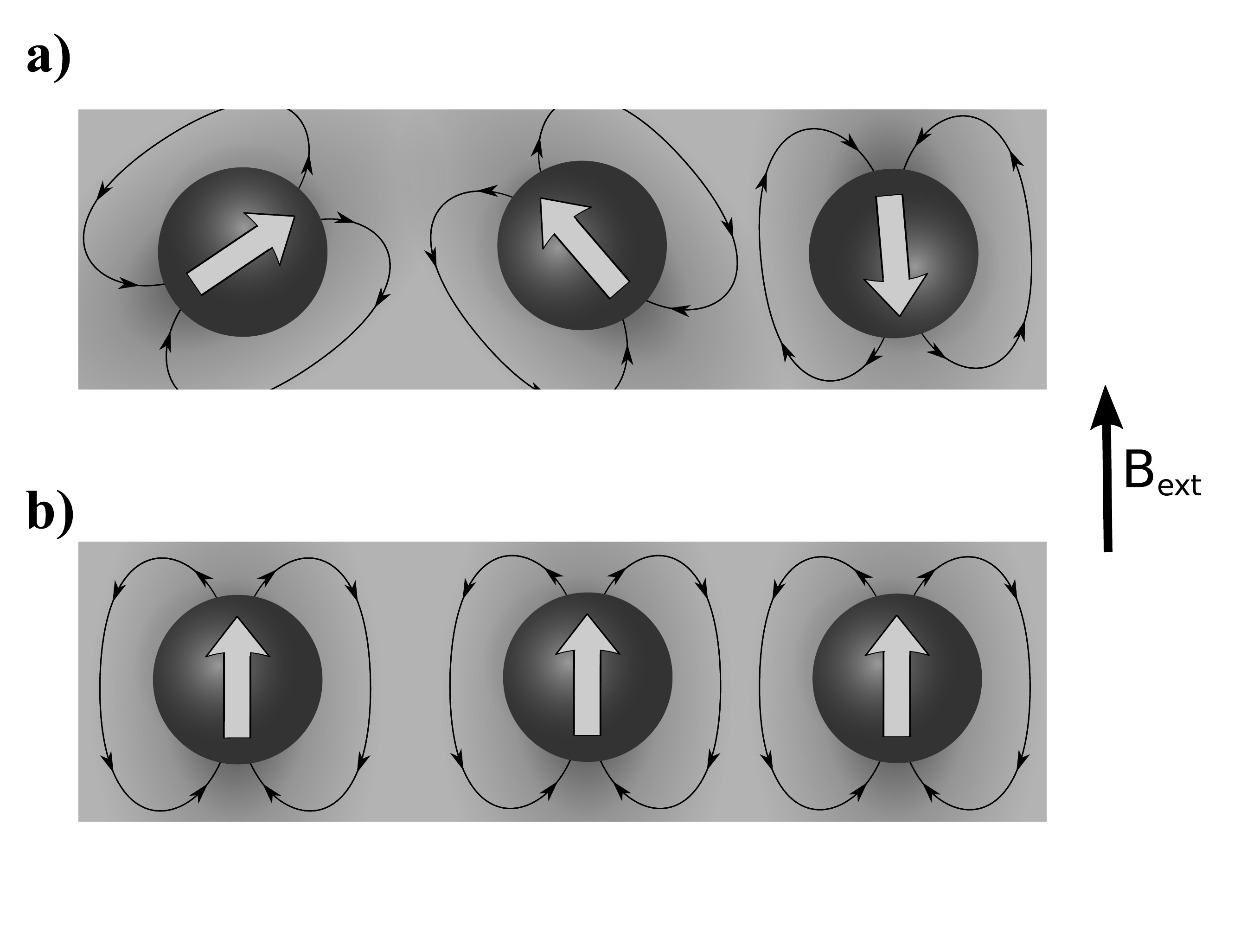}
  \caption{Representation of the magnetic islands in the sample and
    their dipolar fields. \textbf{a)} Randomly oriented islands: The
    dipolar fields average out to zero over the whole
    sample. \textbf{b)} Islands aligned with the applied field: There
    is a net field opposite to the external field in the paramagnetic
    regions.}
  \label{fig:Domains}
\end{figure}

To understand this effect it is important to note that the measured
precessing signal, and thereby the observed field shift, originate
from parts of the sample which are not magnetically ordered.  The
ferromagnetic islands in the sample produce a dipolar field in these
paramagnetic regions. If we assume that the magnetic moment of the
islands are randomly oriented, then their dipolar contributions
average to zero (sketched in Figure~\ref{fig:Domains}a). In such a
case we would expect only an increase in the width of field
distribution sensed by the muons.  Therefore, the shift in the
magnetic field below $T_c$ provides a strong evidence that the
ferromagnetic domains align with the applied field, producing an
average dipolar field in the opposite direction in the paramagnetic
regions (Figure~\ref{fig:Domains}b).  This is consitent with the
reduction of the field shift upon zero-field cooling.  Furthermore, it
agrees with the superparamagnetism, i.e.~weakly coupled magnetically
ordered clusters embedded in a paramagnetic matrix, that has been
reported in Refs.~{[\onlinecite{Lachman2015,Grauer2015}]}.  Both of
these experiments were performed at temperatures below \SI{300}{mK} on
samples with dopant concentrations of $x=0.11$ or $x=0.1$ of V or Cr,
respectively. Based on our results, these samples are not expected to
be fully magnetic.  As the field shift and the paramagnetic volume
fraction decreases with increasing doping level, it may be possible to
suppress this superparamagnetic behavior by increasing the doping
level of the samples.

\subsection{ARPES and Theory}
We observe two peaks in the difference between resonant and off
resonant AIPES. The one close to $E_{\mathrm F}$ is in agreement with
Ref.~[\onlinecite{Peixoto2016}], while the second one has not been
reported before.  The additional increase of intensity at
$E_{\mathrm b}$ below \SI{-4}{eV} (Figure~\ref{fig:ResInt}b) has
previously been attributed to a contribution of a direct Auger
decay~\cite{Peixoto2016}. Here it is less pronounced than in
Ref.~[\onlinecite{Peixoto2016}], possibly due to the lower dopant
concentration in our sample.  The different $h\nu$ values of the two
resonances in the AIPES scan (Figure~\ref{fig:Vres}) point to the
presence of two chemically different V species in the sample.  This
interpretation is consistent with the broader XAS peak that we
observed.

The calculations show that there is a peak in the V-DOS of
substitutional V around $E_{\mathrm b}=$\SIrange{-0.1}{-0.2}{eV},
independent of the presence of additional V in the vdW gap
(Figures~\ref{fig:DiffPlots}e~and~\ref{fig:DFT}b). Therefore, we
attribute the first peak at $E_{\mathrm b}=$\SI{-0.1}{eV} in resonant
AIPES to substitutional V.  In addition, this peak corresponds to a
non-dispersing impurity band in the resonant ARPES measurement, which
is also reproduced by the calculation for substitutional V but absent
for V in the vdW gap, cf. Figure 9c, confirming our conclusion
regarding its origin.

In contrast, the second resonance at higher $h\nu$ and lower binding
energy does not agree with the peaks in our calculations of the V-DOS.
The difference spectra in Figure~\ref{fig:DiffPlots}c features a
general intensity increase around $E_b=\SI{-1.2}{eV}$ and some
additional dispersing features around $\Gamma$ at the same binding
energy. The latter seem to be hybridized with the host band structure
and are likely to be caused by V in the vdW-gap as they agree with the
calculated $\mathbf k$-resolved BSF in Figure~\ref{fig:DFT}c. The
origin of the former is unclear, but we suspect it to be from V that
segregated to the surface of the sample or oxidized during the sample
preparation. In particular, some V oxides are expected to exhibit V
derived states at this binding energy~\cite{Zimmermann1998}.

\subsection{Magnetic coupling mechanism}
Both calculations and resonant ARPES measurements reveal a finite
V-DOS at $E_{\mathrm F}$ and therefore also at $E_{\mathrm b}$
corresponding to the Dirac point, which is expected to be slightly
below $E_{\mathrm F}$~\cite{Zhang2011,Li2016}.  Bulk states at this
binding energy are expected to destroy any quantized transport
signatures, as they open additional conduction channels alongside the
QAH states at the surface~\cite{He2014,Li2016}.  However, they may be
essential to mediate the magnetic coupling at higher temperatures.
Carrier free magnetism has been proposed to either occur via van-Vleck
ferromagnetism or ferromagnetic
superexchange~\cite{Yu2010,Peixoto2016}.  The latter is usually short
ranged and requires a high enough dopant concentration to produce
long-range order.  The site percolation thresholds have been
calculated to be \SI{26.23(2)}{\percent} in a triangular stack lattice
and \SI{69.71(4)}{\percent} for a honeycomb
lattice~\cite{VanDerMarck1997}. These values are much higher than the
magnetic dopant concentration in our samples, where we find a full
volume fraction already at \SI{8}{\percent} substitution of (Bi,Sb).
This implies that a ferromagnetic superexchange alone is not
sufficient to explain the long range magnetic coupling. Note that this
does not exclude the presence of superexchange to trigger nucleation
but requires other longer range interactions for the whole sample to
become ferromagnetic.

The gradual evolution of the magnetic ordering which we see with \mSR\
is another clear indication that there is a broad range of magnetic
interactions involved. This hints at a scenario where charge carrier
mediated interactions dominate.  It would also agree with the finite
V-DOS that we observe at $E_{\mathrm F}$ and with the magnetic moments
on the Sb and Te sites that have been reported by X-ray magnetic
circular dichroism~\cite{Ye2015}.  In addition, DFT calculations on
Bi$_2$Te$_3$ and Sb$_2$Te$_3$ predict the coupling to be caused mainly
by RKKY interactions within the (Bi,Sb) layer and a double exchange
via Te between adjacent layers~\cite{Vergniory2014}.  Both of these
mechanisms rely upon charge carriers at $E_{\mathrm F}$~\cite{Sato2010}.  
Together these findings favor a scenario where
the magnetic interactions are carrier mediated, thus giving a possible
explanation, for the observation of the QAH effect only at dilution
refrigerator temperatures.  It is likely that at these low
temperatures other effects (like disorder induced localization of the
impurity band together with gating) allow to fully gap the system
without affecting the ferromagnetism in the sample.

\section{Conclusion}\label{sec:Concl}
Our \mSR\ measurements on low doping samples indicate a partial volume
fraction of magnetic regions embedded in a paramagnetic
environment. They exhibit no long-range magnetic order even at
temperatures down to \SI{5}{K}.  For higher doped samples
($x \gtrsim 0.16$) we show that the samples become fully magnetic at
low temperatures.  Our findings are consistent with a scenario where
weakly coupled ferromagnetic islands behave as superparamagnetic
clusters which can be aligned with fields as low as \SI{5}{mT}.  Using
resonant SX-ARPES we detect a non-dispersing impurity band close to
$E_{\mathrm F}$ which is doping level independent.  By comparing it
with calculations for two different V sites (V substituting Bi/Sb and
in the vdW gap) we find that this impurity band originates from
substitutional V.  In addition, the calculations for both sites and
the resonant ARPES measurements exhibit a finite V-DOS at
$E_{\mathrm F}$. This implies that the magnetic coupling at high
temperature is predominantly carrier mediated. Along with the partial
magnetic volume fraction, the additional conduction channels
introduced by the impurity band could be another factor limiting the
observation of the QAH effect at elevated temperature.

In this study, we combine for the first time both \lem\ and SX-ARPES
to investigate the same system. These two unique experimental
techniques offer a complementary understanding of the electronic and
magnetic state of the studied materials.  This method of investigation
is generally applicable to any magnetically doped system, including
but not limited to dilute magnetic semiconductors, transition metal
oxides, spin glasses, etc.~and presents a powerful tool to tune and
optimize such materials.  Furthermore this approach can be extended to
magnetic heterostructure and buried interfaces, given that the top
layer is thin enough ($\lesssim\SI{3}{nm}$) in order to allow the
escape of the SX-ARPES photoelectrons.

\section*{Acknowledgments}
We thank Joel Mesot for fruitful discussions, support and valuable
suggestions. We are also grateful to Juraj Krempasky for his advice on
sample preparation for the ARPES measurements. The work at PSI was
supported by the Swiss National Science Foundation (SNF-Grant
No.~200021\_165910). We acknowledge support from DFG through priority
program SPP1666 (Topological Insulators), University of the Basque
Country (Grant Nos. GIC07IT36607 and IT-756-13), Spanish Ministry of
Science and Innovation (Grant Nos.  FIS2013-48286-C02-02-P,
FIS2013-48286-C02-01-P, and FIS2016-75862-P) and Tomsk State
University competitiveness improvement programme (Project No. 8.1.01.2017). 
Partial support by the Saint Petersburg State
University project No. 15.61.202.2015 is also acknowledged. At MIT, C.Z.C. and 
J.S.M. acknowledge support from the STC Center for Integrated Quantum Materials 
under NSF grant DMR-1231319 as well as grants NSF (DMR-1207469, DMR-1700137), 
ONR (N00014-13-1-0301 and N00014-16-1-2657). C.Z.C also acknowledges the support 
from the startup grant provided by Penn State University.
\onecolumngrid
\newpage
\twocolumngrid

\onecolumngrid
\pagebreak
\clearpage
\begin{center}
\section*{\large \textit{Supplemental Material:} Spectroscopic perspective on 
the
interplay between electronic and
magnetic properties of magnetically doped topological insulators}
\twocolumngrid
\end{center}
\setcounter{equation}{0}
\newcounter{FiguresInMainText}
\setcounter{FiguresInMainText}{\value{figure}}
\setcounter{table}{0}
\setcounter{page}{1}
\makeatletter
\renewcommand{\theequation}{S\arabic{equation}}
\renewcommand{\thefigure}{S\the\numexpr\value{figure}-\value{FiguresInMainText}}
\renewcommand{\bibnumfmt}[1]{[S#1]}
\renewcommand{\citenumfont}[1]{S#1}

\section*{Implantation energy dependence of the \mSR\ measurements}
\begin{figure}[htb]
\includegraphics[width=0.9\linewidth]{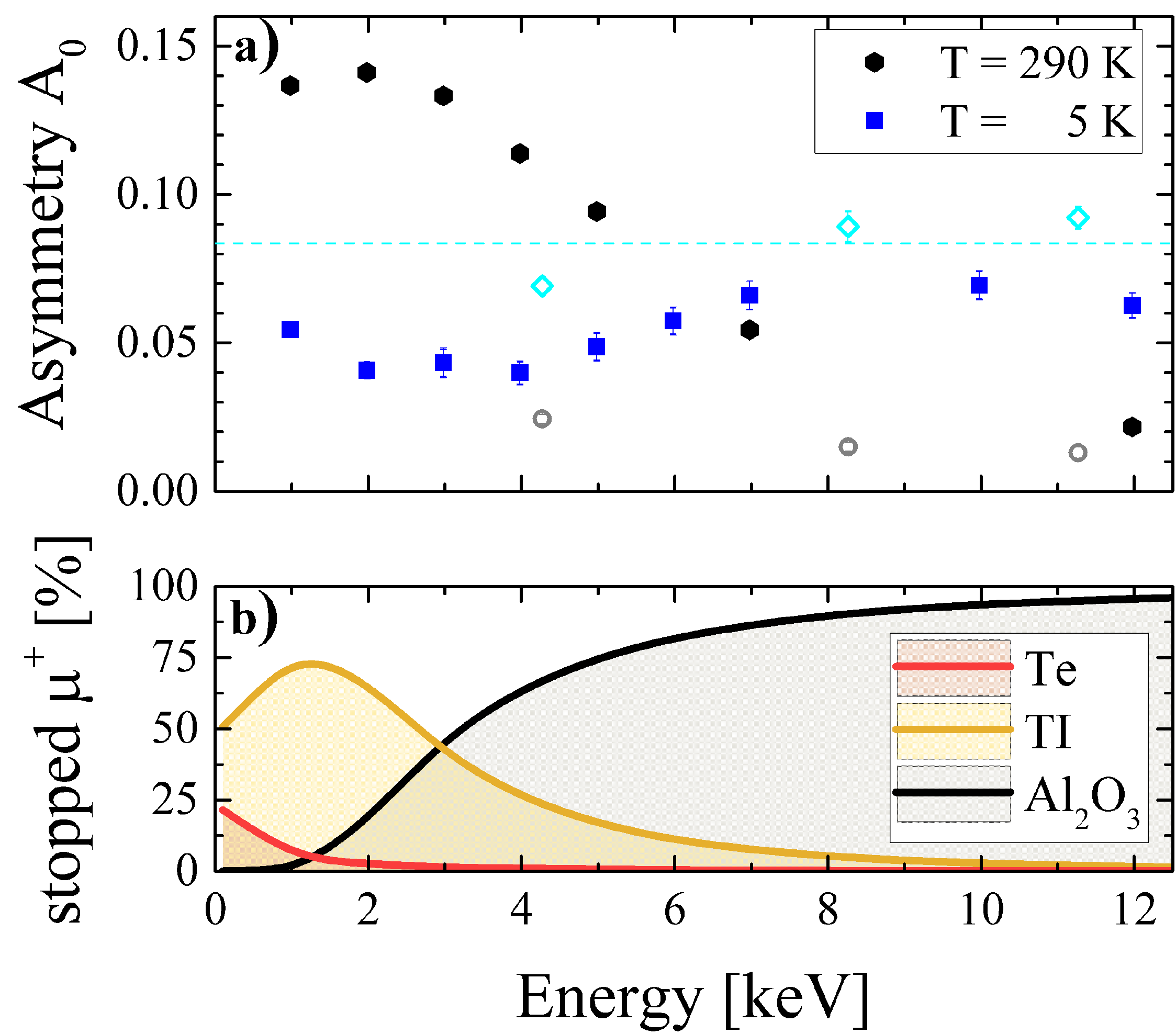}
\vspace{0cm}
\caption{\textbf{a)} Energy dependence of the effective initial
  asymmetry $A_0$ at \SI{290}{K} (black hexagons) and \SI{5}{K} (blue
  squares) measured in the V$_{0.19}$ sample with 5~mT applied
  field. The open symbols show measurements on a bare sapphire
  substrate at \SI{240}{K} (grey circles) and \SI{20}{K} (cyan
  diamonds) with 3~mT applied field. The dashed line shows the average
  value of the asymmetry in sapphire at low temperatures, which was
  used to estimate of the magnetic volume fraction.  \textbf{b)}
  Simulated stopping fraction per layer as a function of implantation
  energy.}
\label{fig:EScan}
\end{figure}

\begin{figure}[htb]
\includegraphics[width=0.9\linewidth]{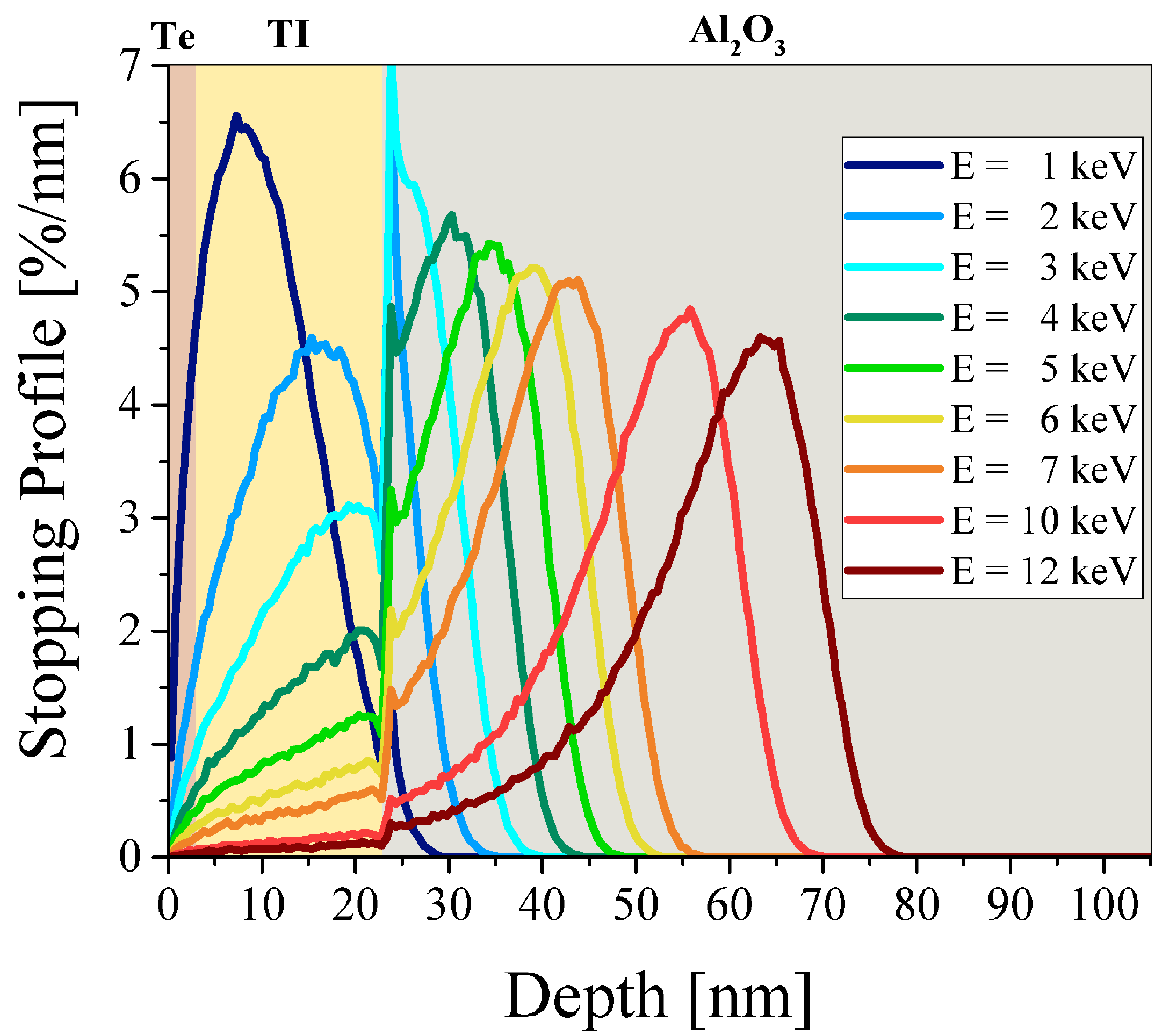}
\vspace{0cm}
\caption{Simulated stopping profiles of the implanted muons as a
  function of depth for different implantation energies.}
\label{fig:profiles}
\end{figure}

Figure~\ref{fig:EScan}a shows the initial asymmetry ($A_0$ in Eq.~1 of
the main text) as a function of implantation energy, together with the
simulated stopping probability in Figure~\ref{fig:EScan}b. The latter
also takes back scattered muons into account, which are relevant at
low implantation energies. The detailed stopping profiles at the
different implantation energies are shown in
Figure~\ref{fig:profiles}. At room temperature and low implantation
energy the wTF-\mSR\ spectra exhibit a high initial asymmetry. With
increasing energy the initial asymmetry first increases, because less
muons are back scattered, and then decreases as more and more muons
are implanted into the sapphire substrate and form muonium, see black
hexagons in Figure~\ref{fig:EScan}a.

At low temperatures the asymmetry measured with high implantation
energy is somewhat increased (blue squares in
Figure~\ref{fig:EScan}a). This effect is observed in all measured
samples and is independent of the exact composition of the TI layer
and is attributed to the temperature dependence of the muonium
formation in sapphire~\cite{brewer2000}. The asymmetry at low
implantation energy is strongly decreased in comparison to its high
temperature value. However, this behavior at low implantation energies
depends strongly on the studied sample, cf.~the measurements at
\SI{2}{keV} shown in Figure~2 in the main text, where the undoped
sample shows almost no change of the initial asymmetry over the full
temperature range. In the magnetic samples, there is a small increase
of the asymmetry towards the lowest implantation energies despite the
enhanced backscattering. This indicates that the stray fields of the
magnetic TI layer are not sufficient to depolarize the muons stopping
in the Te capping.

\begin{figure}[htb]
\includegraphics[width=0.9\linewidth]{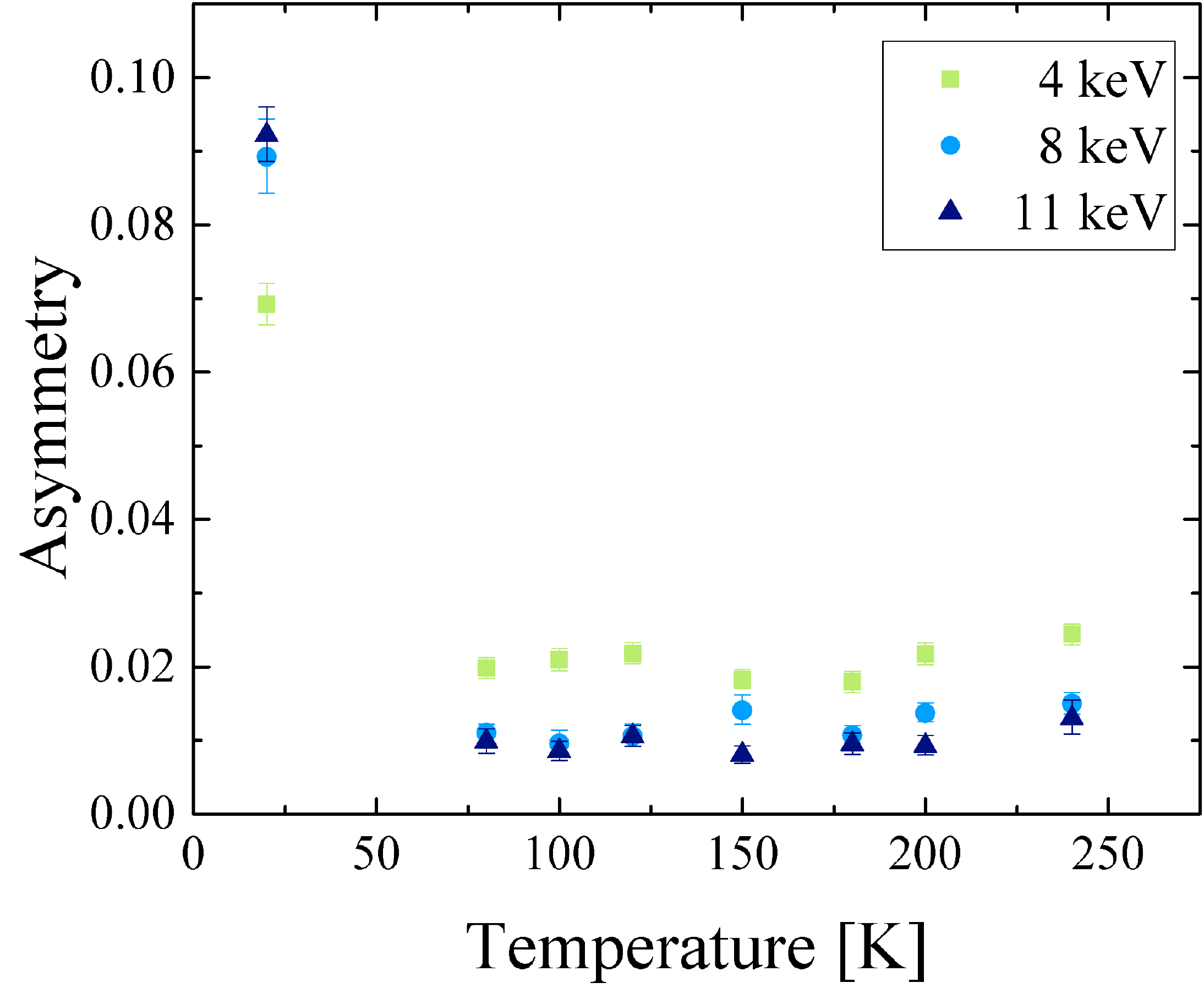}
\vspace{0cm}
\caption{Initial asymmetry in the sapphire substrate as a function of
  temperature at different implantation energies.}
\label{fig:sapphire}
\end{figure}

In addition, the open symbols in Figure~\ref{fig:EScan}a show the
results of reference measurements on a bare sapphire substrate. The
measurements were performed at a temperature of \SI{240}{K} (grey
circles) and \SI{20}{K} (cyan diamonds) in a wTF of \SI{3}{mT}. The
sample used in this measurement was larger than the TI samples and
covered the whole beamspot, and therefore, does not contain any
background contribution from muons landing outside the sample. The
full temperature dependence of the initial asymmetry in the bare
substrate is shown in Figure~\ref{fig:sapphire}, exhibiting a large
increase in asymmetry below $\sim$\SI{50}{K}.

\section*{Calculation of the magnetic volume fraction}
In order to accurately evaluate the magnetic volume fraction of the
doped TI layer, we need to account for the contributions of the
capping layer and substrate to the measured asymmetry. We start by
assuming that muons stopping in the Te capping precess with full
asymmetry while muons stopping in the Ni coated sample backing are
completely depolarized. The fraction of muons stopping in the sapphire
substrate is expected to partly form muonium, and thereby contributing
to the loss of initial asymmetry, while the rest precesses at a
frequency $\omega_{\mathrm{L}}$~\cite{brewer2000}. Our measurements on
the bare substrate show that the initial asymmetry in sapphire is in
general temperature and implantation energy dependent, in agreement
with other reports~\cite{brewer2000,Prokscha2007}. Here we approximate
it with $A_{\mathrm{Al_2O_3}}\approx 0.084\pm0.013$, which is the
average of the measurements on a bare substrate taken at different
implantation energies and $T=$\SI{20}{K}, see dashed line
in~Figure~\ref{fig:EScan}. We also take into account muons landing in
the small exposed areas on the film (where the substrate was clamped
during growth) and assume that they contribute an asymmetry of
$A_{\mathrm{Al_2O_3}}$. This gives us rough estimates for the
contribution to the initial asymmetry from each layer of our samples,
except for the TI.

Using the calculated stopping fraction of muons in each layer
$f_{\mathrm{Layer}}$ (Figures~\ref{fig:EScan}b and
\ref{fig:profiles}), together with the measured initial asymmetry
$A_0$, we can evaluate the contribution of the TI layer to the initial
asymmetry as follows. We start by evaluating the total asymmetry of
the film after correcting for the exposed areas,
\begin{equation}\label{eq:exposed}
  A_{\mathrm{film}}(T)=\frac{1}{1-r}\left(A_0(T)-rA_{\mathrm{
        Al_2O_3 } }\right),
\end{equation}
where $r$ denotes the fraction of the exposed sample area, typically
\SI{5}{\percent}. Then, we determine the contribution to the asymmetry
from muons stopping inside the TI layer using,
\begin{equation}\label{eq:TI}
  A_{\mathrm{TI}}(T)=A_{\mathrm{film}}(T)-f_{\mathrm{Te}}A_{\mathrm{film}}
  (T=\SI{300}{K}) -f_{\mathrm{Al_2O_3}}A_{\mathrm{Al_2O_3}}.
\end{equation}
Finally, the magnetic volume fraction, $f_{\mathrm{MV}}$, in the TI is
given by,
\begin{equation}\label{eq:volfrac}
  f_{\mathrm{MV}}(T)=\frac{ A_{\mathrm{TI}}(T=\SI{300}{K})-
    A_{\mathrm{TI}}(T)}{ A_{\mathrm{TI}}(T=\SI{300}{K})}.
\end{equation}

It is important to point out here that the value of
$A_{\mathrm{Al_2O_3}}$ used in these calculation is the main source of
systematic errors in this calculation. A change in this value will
result in re-scaling the y-axis of Figure~10 in the main text. The
error in $A_{\mathrm{Al_2O_3}}$ can be estimated from the range of
possible values, e.g. between the values measured at high implantation
energy in our samples ($\sim 100\%$ muons stopping in the substrate)
and those reported in Ref.~[\onlinecite{Prokscha2007}] at low
implantation energies. This results in a variation of up to $\sim15\%$
in the volume fraction estimates at the lowest temperature. In this
error estimation we use $A_{\mathrm{Al_2O_3}}$ in the range 0.042 and
0.155, obtained from the temperature average at an implantation
of~\SI{12}{keV} in the V$_{0.19}$ sample (this underestimates the
value at low temperature) and the value from
Ref.~[\onlinecite{Prokscha2007}] at an implantation energy of
\SI{3}{keV} and a temperature of \SI{4}{K}, respectively.

\section*{Field dependence of the mean field}
\begin{figure}[htb]
\includegraphics[width=0.9\linewidth]{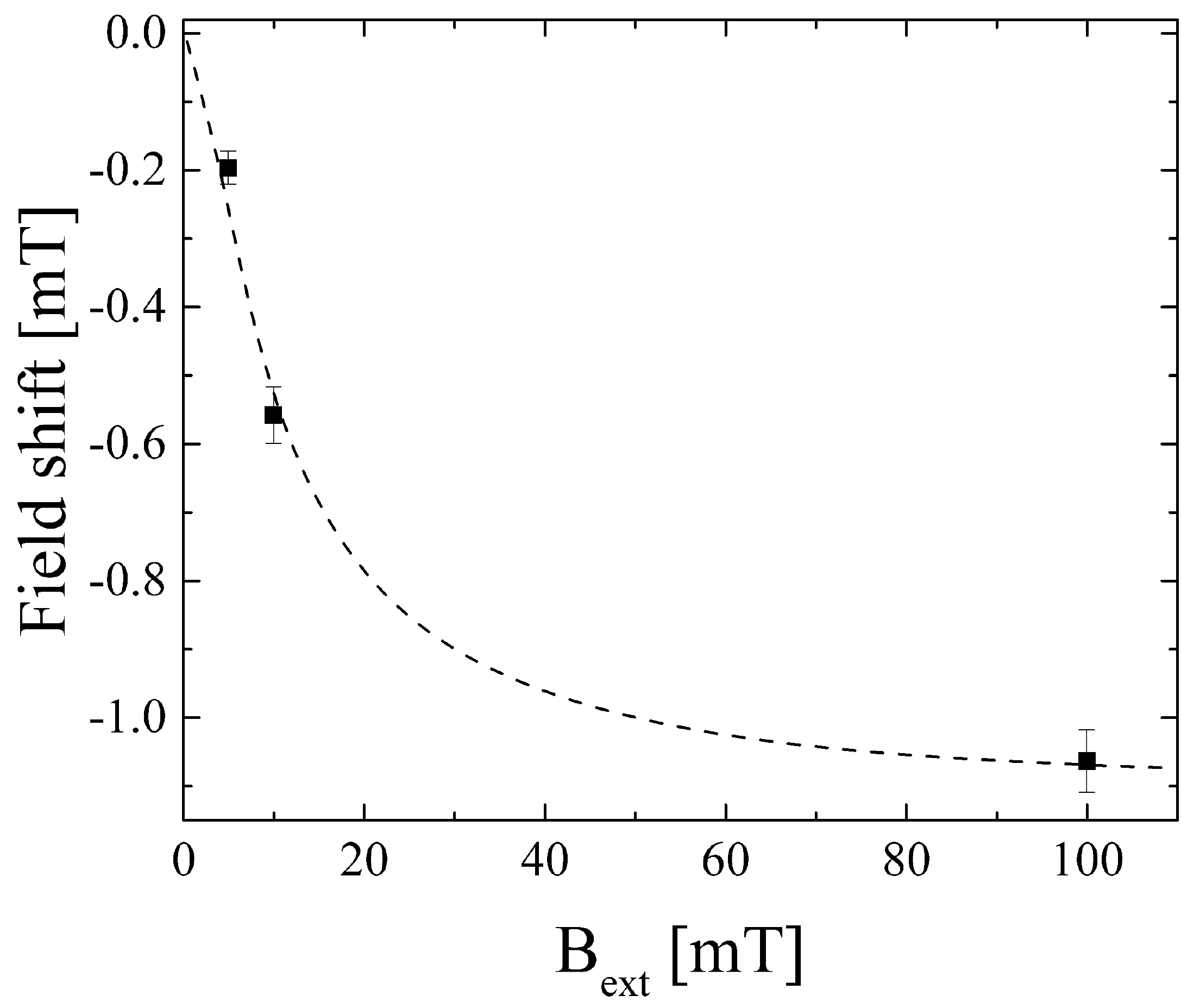}
\vspace{0cm}
\caption{Absolute shift of the mean field between a measurements on
  the V$_{0.08}$ sample at \SI{150}{K} and field cooled at
  \SI{20}{K}. The dashed line is a guide to the eye.}
\label{fig:BScan}
\end{figure}

\begin{figure}[H]
\includegraphics[width=0.9\linewidth]{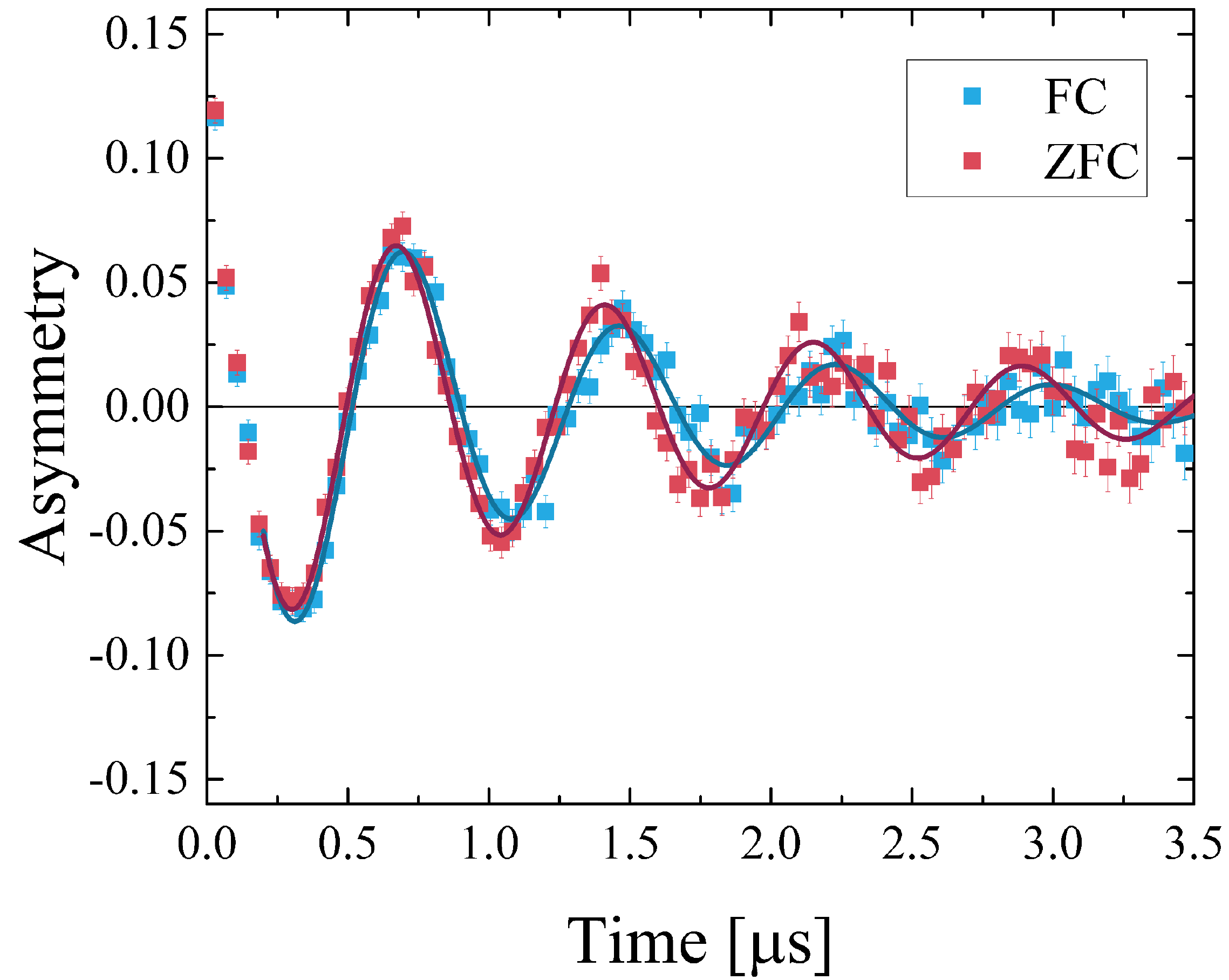}
\caption{Spectra of the V$_{0.08}$ sample at \SI{20}{K} measured after
  zero field cooling and field cooling in an applied field of
  \SI{10}{mT}. There is a clear decrease of the precession frequency
  upon field cooling. The solid lines are fits to Equation~1 of the
  main text, resulting in $B=\SI{9.64(4)}{mT}$ and
  $B=\SI{9.97(3)}{mT}$ for FC and ZFC, respectively.}
\label{fig:ZFC}
\end{figure}

We performed measurements at different fields on the V$_{0.08}$ sample
in order to determine the nature of the field shift we observe at low
temperatures in the wTF-\mSR\ measurements.  We measured the
precession signal at a temperature of \SI{150}{K}, well above the
magnetic transition, and compare it to measurements at \SI{20}{K}. The
shift in the mean field can be extracted from the difference in
precession frequency at \SI{20}{K} and \SI{150}{K}. This is presented
in Figure~\ref{fig:BScan} as a function of applied field.  Note that
the shift is not proportional to the applied field, as one would
expect e.g.~for a Knight shift.
It resembles more the behavior of isolated large magnetic moments in a
magnetic field described by a
Brillouin-function~\cite{Kittel}. Moreover, we find that the field
shift is reduced depending on the cooling history of the sample, as
evident in the difference between the measured spectra after field
cooling (FC) and zero field cooling (ZFC), shown in
Figure~\ref{fig:ZFC}. Together, these results are strong indication
that the magnetic islands in the sample behave like superparamagnets
which are aligned by the small applied field.\\

%

\end{document}